\documentclass[10pt,journal]{IEEEtran}
\usepackage{cite}
\usepackage{amsmath,amssymb,amsfonts}
\usepackage{algorithmic}
\usepackage{algorithm}
\usepackage{graphicx}
\usepackage{textcomp}
\usepackage{makecell}
\usepackage{xcolor}
\usepackage{subfigure}
\usepackage[caption=false,font=normalsize,labelfont=sf,textfont=sf]{subfig}
\usepackage{booktabs}
\usepackage{empheq}
\usepackage{balance}

\newcommand{\revise}[1]{\textcolor{black}{#1}}
\newcommand{\hdel}[1]{}

\begin{document}
\title{DiffPace: Diffusion-based Plug-and-play Augmented Channel Estimation in mmWave and Terahertz Ultra-Massive MIMO Systems\\
}
\author{\IEEEauthorblockN{Zhengdong Hu, Chong Han,~\IEEEmembership{Senior~Member,~IEEE}}, Wolfgang Gerstacker,~\IEEEmembership{Senior~Member,~IEEE}, \\and Robert Schober,~\IEEEmembership{Fellow, IEEE}\thanks{
Z. Hu is with the Terahertz Wireless Communications (TWC) Laboratory, Shanghai Jiao Tong University, Shanghai 200240, China (e-mail: {huzhengdong}@sjtu.edu.cn). 

Chong Han is with the Terahertz Wireless Communications (TWC) Laboratory and also the Cooperative Medianet Innovation Center (CMIC), School of Information Science and Electronic Engineering, Shanghai Jiao Tong University, Shanghai 200240, China (e-mail: chong.han@sjtu.edu.cn).

Wolfgang Gerstacker and Robert Schober are with the Institute for Digital Communications, Friedrich-Alexander-Universität Erlangen-Nürnberg (FAU), 91058 Erlangen,
Germany (e-mail: wolfgang.gerstacker@fau.de, robert.schober@fau.de).
}}
\maketitle
\thispagestyle{empty}
\pdfoptionpdfminorversion=7
\begin{abstract}
Millimeter-wave (mmWave) and Terahertz (THz)-band communications hold great promise in meeting the growing data-rate demands of next-generation wireless networks, offering abundant bandwidth. To mitigate the severe path loss inherent to these high frequencies and reduce hardware costs, ultra-massive multiple-input multiple-output (UM-MIMO) systems with hybrid beamforming architectures can deliver substantial beamforming gains and enhanced spectral efficiency. However, accurate channel estimation (CE) in mmWave and THz UM-MIMO systems is challenging due to high channel dimensionality and compressed observations from a limited number of RF chains, while the hybrid near- and far-field radiation patterns, arising from large array apertures and high carrier frequencies, further complicate CE. \revise{Conventional compressive sensing based frameworks rely on predefined sparsifying matrices, which cannot faithfully capture the hybrid near-field and far-field channel structures, leading to degraded estimation performance.} This paper introduces DiffPace, a diffusion-based plug-and-play method for channel estimation. DiffPace uses a diffusion model (DM) to capture the channel distribution based on the hybrid spherical and planar-wave (HPSM) model. By applying the plug-and-play approach, it leverages the DM as prior knowledge, improving CE accuracy. Moreover, DM performs inference by solving an ordinary differential equation, minimizing the number of required inference steps compared with stochastic sampling method. Experimental results show that DiffPace achieves competitive CE performance, attaining -15 dB normalized mean square error (NMSE) at a signal-to-noise ratio (SNR) of 10 dB, with 90\% fewer inference steps compared to state-of-the-art schemes, simultaneously providing high estimation precision and enhanced computational efficiency.
\end{abstract}

\section{Introduction}
Millimeter-wave (mmWave) and Terahertz (THz)-band communications have emerged as transformative technologies for next-generation wireless networks, leveraging abundant bandwidth to meet unprecedented data rate demands~\cite{b10,a1}. Specifically, these technologies provide promising solutions to overcome spectrum scarcity and capacity limitations in existing wireless systems, enabling long-awaited applications. These applications range from the metaverse and extended reality (XR) to ultra-fast wireless fronthaul and backhaul, as well as the integration of millimeter-level sensing with terabit-per-second (Tbps) communications. Moreover, mmWave and THz communications hold great potential for advancing edge intelligence, high-speed data centers, and inter-satellite links, paving the way for ultra-reliable and high-capacity wireless connectivity~\cite{a1}. 

However, these high-frequency bands face unique propagation challenges, including high spreading loss and molecular absorption, which limit communication distances~\cite{channel_survey}. To mitigate these limitations, ultra-massive multiple-input multiple-output (UM-MIMO) systems have emerged as a promising solution, offering high beamforming gains by leveraging arrays with hundreds to thousands of antennas. To reduce hardware costs, hybrid beamforming architectures are employed in UM-MIMO, enabling the control of a large number of antennas with a limited number of RF chains~\cite{b8,b9}. However, these systems rely on accurate channel state information (CSI) for effective hybrid precoding and combining, which necessitates reliable and precise channel estimation (CE).

In hybrid UM-MIMO systems, CE requires recovering high-dimensional channel matrices from a limited number of pilot observations, making it a highly ill-posed inverse problem. Additionally, the hybrid near- and far-field radiation patterns, induced by large array apertures and high carrier frequencies, further complicate CE. As a result, achieving accurate CE in hybrid UM-MIMO systems operating across near- and far-field regions remains a critical but challenging task, driving the need for advanced CE techniques tailored to cross-field scenarios.

\subsection{Related Work}
In the literature, CE methods can be broadly classified into conventional approaches\cite{b14,b13,ce_hybrid_lcomm2021,omp11,new6,ce_hybrid_tcomm2023,a2,b27,b12,ce1_cunhua,editor3_trace}, and deep learning (DL) based methods\cite{ce_cnn_jstsp2019,ce_cnn_tcom2021,b22,ce_fpn_jstsp2023,b17,ce_gan_jsac2021,ce_dm_twc2023,b30,b15,b21,b24,b6,editor2_dlce,editor4_dlce,editor6_dl}. Conventional methods are typically divided into on-grid and off-grid approaches. On-grid methods assume that channel parameters, such as angles of arrival and departure, lie on a predefined grid. A prominent class of these methods includes compressive sensing (CS) techniques, such as the orthogonal matching pursuit (OMP)\cite{b14} and approximate message passing (AMP)\cite{b13}. These methods exploit the sparse nature of the channel by constructing a dictionary matrix and estimating a sparse representation, thereby enabling CE with reduced pilot overhead. For cross-field scenarios, a two-stage CS algorithm is proposed in~\cite{ce_hybrid_lcomm2021}, which exploits the sparsity of both the far-field and near-field paths in the angular and polar domains. Specifically, OMP is firstly used with an angular-domain codebook to estimate the far-field components, followed by a polar-domain codebook to detect the near-field paths. In~\cite{omp11}, a simultaneous OMP algorithm in the polar domain is introduced for efficient near-field CE. Furthermore, a sparse message passing (SMP) algorithm is developed in~\cite{new6}, which iteratively detects nonzero entries in the sparse channel matrix. However, the assumption of sparsity does not always hold in practical propagation environments, especially in near-field scenarios. In such cases, sparsity alone is insufficient to model real-world channel structures. Additionally, the performance of CS-based methods is often limited by the grid mismatch problem, as the true channel parameters do not exactly align with the predefined grids.
To mitigate these issues, off-grid methods have been proposed. These methods avoid fixed grid assumptions and instead refine the grid resolution iteratively to improve estimation accuracy. For example, the authors of~\cite{ce_hybrid_tcomm2023} first estimate the line-of-sight paths using an on-grid method and then apply gradient descent for iterative refinement. \hdel{In~\cite{a2,b27}, multiple signal classification (MUSIC) algorithm is employed to exploit channel subspace information through eigenvalue decomposition.}Although off-grid methods typically outperform on-grid ones, they often suffer from high computational complexity due to their iterative nature.

With the rapid development of DL, DL-based methods have attracted significant attention for CE. These methods can learn the complex structures inherent to wireless channels and are generally categorized into data-driven and model-driven approaches. Data-driven methods train neural networks to directly map the received pilot signals to the full channel matrix or its parameters. The authors of~\cite{ce_cnn_jstsp2019} design a deep convolutional neural network (CNN) to estimate the channel matrix by learning spatial correlations. Similarly, the authors of~\cite{ce_cnn_tcom2021} employ a CNN to estimate channel parameters. In~\cite{b22}, a deep image prior (DIP) network is used to denoise the received signal, followed by least-squares-based channel recovery. While data-driven methods can achieve high CE accuracy, they often function as ``black boxes", lacking interpretability and theoretical guarantees on generalization capabilities.

In contrast, model-driven methods incorporate domain knowledge into the network architecture, typically by unfolding traditional iterative algorithms into trainable network layers. This approach reduces the number of trainable parameters and improves interpretability. In~\cite{ce_fpn_jstsp2023}, the orthogonal approximate message passing (OAMP) algorithm is embedded into a deep neural network using a fixed-point structure. Similarly, the learned denoising AMP method in~\cite{b17} replaces the shrinkage function in AMP with a trainable denoising network. While these unfolded networks improve performance and interpretability, they are still constrained by the limitations of the underlying algorithms. Moreover, their dependence on a fixed measurement matrix restricts their adaptability to changes in pilot configurations or measurement conditions.

Beyond these approaches, the success of large language models has inspired recent advances in generative AI~\cite{llm1_cunhua}, opening new avenues for model-driven CE. Deep generative models have demonstrated remarkable capabilities in modeling complex, high-dimensional data distributions, making them particularly suitable for CE tasks. The work in \cite{ce_gan_jsac2021} employs a generative adversarial network (GAN) within a compressive sensing framework, where the GAN learns a low-dimensional latent representation of the channel and gradient descent projects measurements onto this learned manifold. As the model training is decoupled from the measurement process, it achieves strong generalization across varying pilot configurations. Notably, this approach eliminates dependence on predefined statistical channel models by directly learning the distribution, demonstrating robust performance even in out-of-distribution scenarios where test channels statistically differ from training data. 

Despite these advantages, GANs suffer from well-known training instability problems inherent to their adversarial optimization framework. This limitation has spurred the adoption of more stable alternatives such as diffusion models, which have outperformed GANs in multiple domains \cite{diffusion_beat_gan}. Their application to CE is demonstrated in \cite{ce_dm_twc2023}, where a score-based generative diffusion model learns the MIMO channel distribution, enabling CE through posterior sampling. While achieving impressive performance and excellent generalization, this state-of-the-art approach faces practical challenges, as the inference process typically requires hundreds to thousands of iterative steps for stochastic sampling, potentially hindering real-time deployment. Furthermore, the work~\cite{ce_dm_twc2023} focuses exclusively on far-field scenarios, leaving near-field and cross-field effects prevalent in mmWave and THz communications unexplored.
\begin{figure*}[t]
    \centering
 \includegraphics[width=1.0\textwidth]{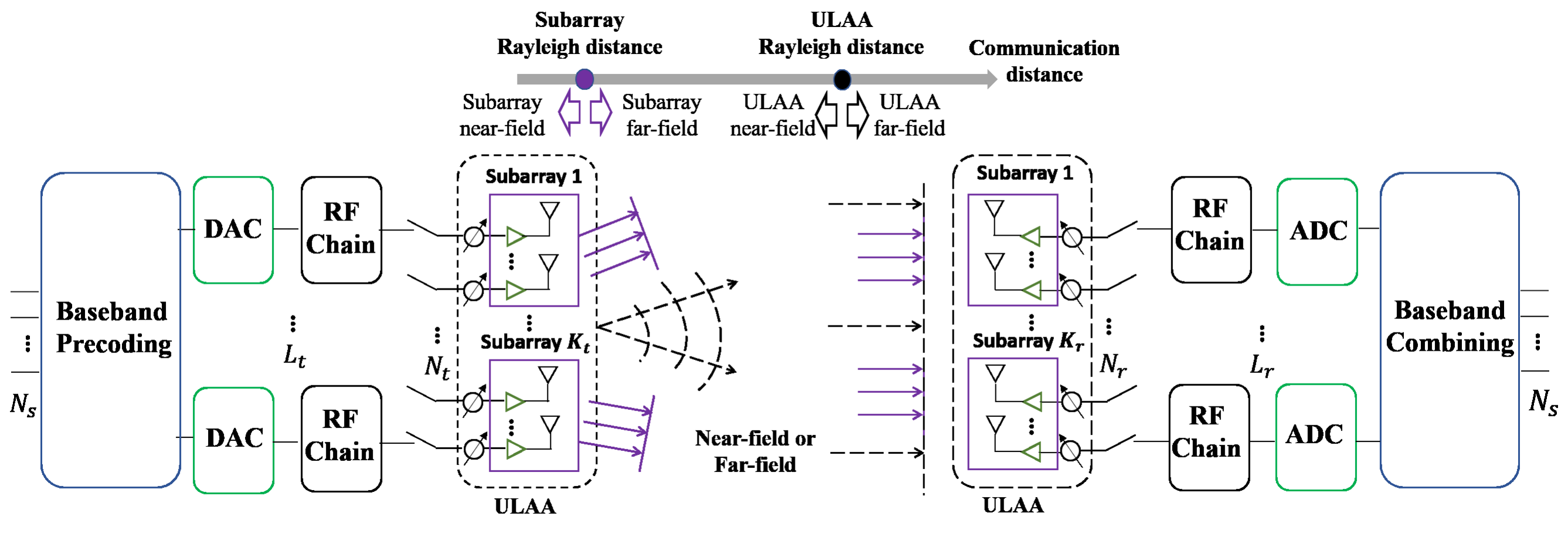} 
    \caption{MmWave and THz UM-MIMO systems with hybrid analog-digital precoding and combining.}
    \label{fig_system}
\end{figure*}

\subsection{Contributions}

In this paper, we propose an efficient plug-and-play augmented diffusion model for CE, named DiffPace, tailored for cross-field scenarios in mmWave and THz UM-MIMO systems. These systems simultaneously exhibit both near-field and far-field propagation characteristics, which pose unique challenges for unified CE. To address this, we train the diffusion model using data generated from the hybrid planar-spherical wave model (HPSM)~\cite{hpsm_yuhan}. In this model, planar wavefronts are assumed within each subarray, while spherical wavefronts are used for inter-subarray propagation. Leveraging this realistic modeling framework enables the diffusion model to generalize well across both the near-field and far-field conditions without requiring separate estimation strategies, thus supporting a unified estimation pipeline. DiffPace utilizes a diffusion model to learn the complex distributions of realistic wireless channels, without relying on strong assumptions such as sparsity or a pre-defined structure. This generative prior enables robust CE in challenging and diverse environments.

To enhance inference efficiency, we further propose a plug-and-play ordinary differential equation (ODE) based solver that significantly reduces the number of denoising steps required during sampling. This not only accelerates the estimation process, but also allows the pre-trained diffusion model to flexibly adapt to various system configurations, such as different pilot lengths or signal-to-noise ratios, without the need for retraining. Extensive simulation results validate the effectiveness of DiffPace, demonstrating its superior estimation accuracy and computational efficiency compared to conventional and DL-based methods.

The key contributions of this work are summarized as follows:

\begin{itemize} \item \textbf{Diffusion-Based Channel Modeling:} We propose a diffusion model-based channel modeling framework that learns powerful prior knowledge over HPSM channel models, which generalizes across near-field and far-field conditions within a single unified framework.


\item \textbf{Plug-and-Play Optimization with ODE Solver:} We design an ODE-based sampling method integrated with a plug-and-play strategy, enabling efficient and accurate channel recovery with significantly fewer inference steps compared to state-of-the-art schemes.

\item \textbf{Comprehensive Evaluation:} Simulation results confirm that DiffPace achieves high accuracy and strong generalization across diverse channel conditions, making it highly practical for real-time wireless communication systems. \end{itemize}

The remainder of this paper is organized as follows. The system model for the mmWave and THz UM-MIMO and the CE problem are established in Section~\ref{sec:system model}. The DiffPace method is elaborated in Section~\ref{sec:method}. Section~\ref{sec:evaluation} evaluates the performance of the proposed methods. Finally, the conclusion is drawn in Section~\ref{sec:conclusion}.

\textbf{Notation:} 
$a$ is a scalar. \textbf{a} denotes a vector. 
\textbf{A} represents a matrix.
$(\cdot)^T$ denotes the transpose.
$(\cdot)^*$ refers to the conjugate transpose.
$E\{\cdot \}$ stands for the expectation.
$ \left\|\cdot\right\|_1 $, $ \left\|\cdot\right\|_2 $, and $ \left\|\cdot\right\|_F $ represent the L1 norm, L2 norm, and Frobenius norm, respectively.
${\rm vec}\{\cdot \}$ represents the vectorization of a matrix.
$|\cdot|$ denotes the absolute value. $\ast$ is the convolution operation. $\textbf{I}_N$ is the $N$ dimensional identity
matrix. $\otimes$ refers to the Kronecker product.

\section{System Model and Problem Formulation}\label{sec:system model}
In this section, we first present the system model for mmWave and THz UM-MIMO communication systems. We then introduce the HPSM, which provides a unified framework encompassing both far-field and near-field propagation characteristics. The HPSM enables a sparse angular representation of the channel which significantly reduces the complexity of CE. Building upon these foundations, we formulate the CE problem by exploiting the inherent sparsity of the HPSM representation to develop efficient estimation algorithms.

\subsection{System Model}
\revise{In this work, we consider a widely spaced multi-subarray (WSMS) UM-MIMO architecture for mmWave and THz communications. The WSMS structure employs multiple subarrays with enlarged inter-subarray spacing, which enhances the spatial multiplexing capability and significantly improves the spectral efficiency of mmWave and THz systems~\cite{WSMS}. Based on this architecture, we study an orthogonal frequency division multiplexing (OFDM) based system that utilizes hybrid precoding and combining at both transmitter (Tx) and receiver (Rx)~\cite{b12}.} While the system operates using OFDM, our analysis focuses on CE for a single subcarrier, and thus we omit explicit subcarrier indexing in the following discussions.

As shown in Fig.~\ref{fig_system}, the system employs ultra-large antenna arrays (ULAAs) consisting of \(N_t\) transmit and \(N_r\) receive antennas at Tx and Rx, respectively. These ULAAs are organized into \(K_t\) and \(K_r\) distinct subarrays at Tx and Rx, enabling efficient signal processing. \revise{Within each subarray, the antenna elements are spaced by half wavelength, whereas the spacing between subarrays is multiple wavelengths~\cite{WSMS}.} Moreover, there are $L_t$ and $L_r$ RF chains to connect with the antennas at Tx and Rx, respectively. The number of data streams $N_s$ satisfies $N_s\leq L_t < N_t$ and $N_s\leq L_r < N_r$ \cite{b12}. The transmitted symbol vector $\mathbf{s}\in\mathbb{C} ^{N_s \times 1}$ is precoded to form the transmitted complex signal $\mathbf{x}_c\in\mathbb{C} ^{N_t \times 1}$ as
\begin{equation}
    \mathbf{x}_c = \mathbf{F}_{\mathrm{RF}}\mathbf{F}_{\mathrm{BB}}\mathbf{s}, \label{e1}
\end{equation}
where the analog precoder $\mathbf{F}_{\mathrm{RF}}\in \mathbb{C}^{N_t\times L_t}$ is implemented using phase shifters,with each element satisfying the constant modulus constraint. Specifically, $\mathbf{F}_{\mathrm{RF}}[i,n] = \frac{1}{\sqrt{N_t}}{e}^{j f_{i,n}}$, where $f_{i,n}\in [0,2\pi]$ represents the phase shift value, and $i,n$ represents the element position. Additionally, $\mathbf{F}_{\mathrm{BB}}\in \mathbb{C}^{L_t \times N_s}$ represents the baseband digital precoder. After traversing the channel, the received signal at Rx is processed by the analog combiner \(\mathbf{W}_{\mathrm{RF}} \in \mathbb{C}^{N_r \times L_r}\) and digital combiner \(\mathbf{W}_{\mathrm{BB}} \in \mathbb{C}^{L_r \times N_s}\), yielding the baseband signal vector \(\mathbf{y}_c \in \mathbb{C}^{N_s \times 1}\). This process can be represented by
\begin{equation}
\mathbf{y}_c=\mathbf{W}_{\mathrm{BB}}^*\mathbf{W}_{\mathrm{RF}}^*\mathbf{H}\mathbf{x}+\mathbf{W}_{\mathrm{BB}}^*\mathbf{W}_{\mathrm{RF}}^*\mathbf{n}_c, \label{eq4}
\end{equation}
where \(\mathbf{H} \in \mathbb{C}^{N_r \times N_t}\) is the frequency-domain channel matrix, and \(\mathbf{n}_c\sim \mathcal{CN}(0, \sigma^2_n \mathbf{I}_{N_r})\) is the additive noise vector with noise power \(\sigma^2_n\). The analog combiner \(\mathbf{W}_{\mathrm{RF}}\), implemented through phase shifters, satisfies the constant modulus constraint. By incorporating \eqref{e1} and \eqref{eq4}, the received signal can be represented by
\begin{equation}
\mathbf{y}_c = \overline{\mathbf{W}}^{*} \mathbf{H} \overline{\mathbf{F}} \mathbf{s} + \overline{\mathbf{n}}_c,
\end{equation}
where \(\overline{\mathbf{W}} = \mathbf{W}_{\mathrm{RF}} \mathbf{W}_{\mathrm{BB}} \in \mathbb{C}^{N_r \times L_r}\) denotes the cascaded combiner, \(\overline{\mathbf{F}}\ = \mathbf{F}_{\mathrm{RF}} \mathbf{F}_{\mathrm{BB}} \in \mathbb{C}^{N_t \times L_t}\) represents the cascaded precoder, and the noise vector \(\overline{\mathbf{n}}_c=\overline{\mathbf{W}}^*\mathbf{n}_c \in \mathbb{C}^{L_r \times 1}\) represents the combined noise at the receiver. The parameters of the mmWave and THz UM-MIMO system are listed in Table~\ref{tab:parameter}.

\begin{table}[t]
\caption{Parameters of the considered mmWave and THz UM-MIMO systems}
\begin{center}
\begin{tabular}{cc}
\bottomrule
\textbf{Parameter}& \textbf{Symbol} \\
\midrule
Number of antennas at Tx & $N_t$ \\
Number of antennas at Rx & $N_r$ \\
Number of transmitted data streams & $N_s$\\
Number of subarrays at Tx & $K_t$\\
Number of subarrays at Rx & $K_r$\\
Number of RF chains at Tx & $L_t$\\
Number of RF chains at Rx & $L_r$\\
Transmitted symbol at Tx & $\mathbf{s}$\\
Transmitted signal at Tx & $\mathbf{x}_c$\\
Received signal at Rx & $\mathbf{y}_c$\\
Analog precoder at Tx & $\mathbf{F}_{\mathrm{RF}}$ \\
Digital precoder at Tx & $\mathbf{F}_{\mathrm{BB}}$\\
Training precoder at Tx & $\mathbf{F}_{\mathrm{t}}$\\
Analog combiner at Rx &  $\mathbf{W}_{\mathrm{RF}}$ \\
Digital combiner at Rx & $\mathbf{W}_{\mathrm{BB}}$\\
Training combiner at Rx &  $\mathbf{W}_{\mathrm{r}}$\\
Noise vector at Rx & $\mathbf{n}_c$\\
Channel matrix & $\mathbf{H}$\\
Number of paths & $L$\\
Complex gain of the $l^{\mathrm{th}}$ path & $\alpha_l$\\
Delay of the $l^{\mathrm{th}}$ path & $\tau_l$\\
Angle pair of departure for the $l^{\mathrm{th}}$ path & ($\theta_{T,l},\phi_{T,l}$)\\
Angle pair of arrival for the $l^{\mathrm{th}}$ path & ($\theta_{R,l},\phi_{R,l}$)\\
Array steering vector & $\mathbf{a}$\\
Number of training frames & $M$\\
Measurement matrix &$\boldsymbol{\Phi}$\\
Vectorized channel vector & $\mathbf{h}$\\
\bottomrule
\end{tabular}
\label{tab:parameter}
\end{center}
\end{table}

\subsection{HPSM Channel Model}
In mmWave and THz UM-MIMO systems, the combination effects of large ULAA apertures and short operating wavelengths lead to coverage distances spanning both the near-field and far-field regions. The boundary between these regions is defined by the Rayleigh distance, which grows quadratically with array size and inversely with wavelength~\cite{rayleigh_add}. As illustrated in Fig.~\ref{fig_system}, the ULAA of Rx may lie in either the near field or far field of the Tx's ULAA depending on their separation distance. To properly characterize such systems, we first analyze the distinct channel models for both propagation regimes, then introduce the HPSM channel model that unifies near-field and far-field behavior within a single framework.

In near-field UM-MIMO systems, the channel can be characterized by the spherical wave model (SWM). The channel response between the $i^{\mathrm{th}}$ transmit antenna and the $n^{\mathrm{th}}$ receive antenna is expressed as~\cite{pwm_swm}
\begin{equation}
    \mathbf{H}_{S}[i,n]=\sum_{l=1}^{L}|\alpha^{i,n}_l|e^{-j\frac{2\pi }{\lambda}d^{i,n}_l},
\end{equation}
where $l = 1, \dots, L$ denotes the index of propagation paths between the $i^{\mathrm{th}}$ Tx and $n^{\mathrm{th}}$ Rx antennas, $|\alpha^{i,n}_l|$ represents the magnitude of the path gain for the $l^{\mathrm{th}}$ path, and $d^{i,n}_l$ is the corresponding propagation distance. The phase term $-j\frac{2\pi d^{i,n}_l}{\lambda}$ captures the phase shift associated with the complex path gain $\alpha^{i,n}_l$. For each antenna pair $(i, n)$, the SWM requires $2L$ parameters $\{|\alpha^{i,n}_l|, d^{i,n}_l\}$ to fully describe the channel response. Consequently, the total number of parameters in the complete Tx-Rx channel matrix scales as $2LN_rN_t$, which becomes prohibitively large for mmWave and THz UM-MIMO systems due to their extreme array dimensions~\cite{hpsm_yuhan}.

In the far field, the channel can be modeled using the planar wave model (PWM), which serves as a simplified approximation of the SWM when the array aperture is significantly smaller than the communication distance~\cite{pwm_swm}. Under this assumption, the propagating waves follow parallel paths with planar wavefronts. For mmWave and THz UM-MIMO systems, the channel matrix under the PWM is given by~\cite{pwm_swm}
\begin{equation}
 \mathbf{H}_{P} = \sum_{l=1}^L \alpha_l e^{-j\phi_l} \mathbf{a}_R(\theta_{R,l}, \phi_{R,l}) \mathbf{a}_T^H(\theta_{T,l}, \phi_{T,l}),   
\end{equation}
where $L$ represents the number of propagation paths, $\alpha_l$ denotes the complex gain of the $l^{\mathrm{th}}$ path, and $\phi_l$ corresponds to the phase shift. The array steering vectors $\mathbf{a}_R(\theta_{R,l}, \phi_{R,l}) \in \mathbb{C}^{N_r \times 1}$ at Rx and $\mathbf{a}_T(\theta_{T,l}, \phi_{T,l}) \in \mathbb{C}^{N_t \times 1}$ (transmitter) at Tx are characterized by their respective angles of arrival (AoA) $(\theta_{R,l}, \phi_{R,l})$ and departure (AoD) $(\theta_{T,l}, \phi_{T,l})$.
For a uniform planar array (UPA) placed in the $xz$-plane of a 3D Cartesian coordinate system, with $N_x$ and $N_z$ antenna elements uniformly spaced along the $x$- and $z$-axes respectively, the array steering vector is given by
\begin{equation}
\mathbf{a}(\theta, \phi) = \frac{1}{\sqrt{N_x N_z}} \left( e^{-j\pi \sin\theta \cos\phi\, \mathbf{n}_x} \otimes e^{-j\pi \sin\phi\, \mathbf{n}_z} \right),
\end{equation}
where $\mathbf{n}_x = [0, 1, \dots, N_x-1]^T$ and $\mathbf{n}_z = [0, 1, \dots, N_z-1]^T$ denote the antenna element indices along the $x$- and $z$-axes, respectively.
The PWM requires $6L$ parameters $\{\alpha_l, \phi_l, \theta_{R,l}, \phi_{R,l}, \theta_{T,l}, \phi_{T,l}\}$ that are independent of the array dimensions, making it significantly more parameter-efficient than the SWM~\cite{hpsm_yuhan}. However, this computational advantage comes at the cost of reduced accuracy in near-field scenarios, where the planar wave assumption becomes invalid.

To achieve an optimal trade-off between modeling accuracy and computational complexity, we employ the HPSM~\cite{hpsm_yuhan} for the transitional region spanning near-field to far-field operation. As illustrated in Fig.~\ref{fig_system}, the limited aperture of subarrays leads to small Rayleigh distances, ensuring that each Rx subarray operates in the far-field region of every Tx subarray. In contrast, the entire ULAA exhibits a substantially larger Rayleigh distance, meaning the entire Rx ULAA may reside in either the near-field or far-field region of the entire Tx ULAA. Consequently, we model the ULAA-level interactions using the SWM to maintain accuracy, while adopting the PWM for intra-subarray propagation. The HPSM framework leverages the fact that subarrays share common reflectors in the propagation environment, resulting in the same number of paths across subarrays, while exhibiting distinct amplitudes, phases, AoA, and AoD according to the SWM principles. The channel response between the $k_t^{\mathrm{th}}$ Tx subarray and the $k_r^{\mathrm{th}}$ Rx subarray is then given by
\begin{equation}\label{hpsm}
\begin{aligned}
    \mathbf{H}_{HPSM}[k_t,k_r]=&\sum_{l=1}^L|a_l^{k_t,k_r}|e^{-j\phi_l^{k_t,k_r}}\\
    &\mathbf{a}_{k_r}(\theta_{k_r,l}, \phi_{k_r,l})
    \mathbf{a}_{k_t}^H(\theta_{k_t,l}, \phi_{k_t,l}),
\end{aligned}
\end{equation}
where $|a_l^{k_t,k_r}|$ denotes the $l^{\mathrm{th}}$ path gain magnitude between the $k_t^{\mathrm{th}}$ Tx subarray and $k_r^{\mathrm{th}}$ Rx subarray, $\phi_l^{k_t,k_r}$ represents the subarray-specific phase shift, and $\mathbf{a}_{k_r}(\theta_{k_r,l}, \phi_{k_r,l}) \in \mathbb{C}^{N_r \times 1}$ and $\mathbf{a}_{k_t}(\theta_{k_t,l}, \phi_{k_t,l}) \in \mathbb{C}^{N_t \times 1}$ are the steering vectors for the Rx and Tx subarrays, respectively. 

The parameter set $\{|a_l^{k_t,k_r}|,\phi_l^{k_t,k_r}, \theta_{k_r,l}, \phi_{k_r,l}, \theta_{k_t,l}, \phi_{k_t,l}\}$ includes only $6LK_tK_r$ parameters, offering significant complexity reduction compared to the pure SWM while maintaining accuracy. This hybrid approach effectively balances computational efficiency with CE precision, enabling seamless operation across both the near-field and far-field regimes. The HPSM thus provides a smooth transition between propagation domains while maintaining model fidelity.

\subsection{Sparse Angular Representation}
In the far field, the channel matrix \(\mathbf{H}_P\) can be represented in the angular domain as
\begin{equation}
\mathbf{H}_P = \mathbf{A}_R \mathbf{H}_b \mathbf{A}_T^*,
\end{equation}\label{eq_beamrep}
where \(\mathbf{A}_R \in \mathbb{C}^{N_r \times N_r}\) and \(\mathbf{A}_T \in \mathbb{C}^{N_t \times N_t}\) are the 2D Discrete Fourier Transform (DFT) matrices for the array responses at the Tx and Rx, respectively, and \(\mathbf{H}_b \in \mathbb{C}^{N_r \times N_t}\) is the beamspace channel matrix. This representation is crucial for exploiting channel sparsity.

The same principle applies to the HPSM~\cite{hpsm_yuhan}. In this model, the subarray channel responses follow the PWM form, where each Rx subarray contains \( N_r^{\text{sub}} = N_r / K_r \) antennas and each Tx subarray contains \( N_t^{\text{sub}} = N_t / K_t \) antennas. We assume that \( N_r \) and \( N_t \) are divisible by \( K_r \) and \( K_t \), respectively. Under this setting, the subarray-based codebooks are given by \( \mathbf{A}_{k_r} \in \mathbb{C}^{N_r^{\text{sub}} \times N_r^{\text{sub}}} \) at the Rx subarray and \( \mathbf{A}_{k_t} \in \mathbb{C}^{N_t^{\text{sub}} \times N_t^{\text{sub}}} \) at the Tx subarray.
 For the Rx ULAA, the codebook is constructed as
\begin{equation}
\overline{\mathbf{A}}_R = \text{blkdiag} \left[ \mathbf{A}_{1}, \ldots, \mathbf{A}_{K_r} \right].
\end{equation}
The Tx codebook \(\overline{\mathbf{A}}_T \in \mathbb{C}^{N_t \times N_t}\) has a similar formulation. Therefore, the HSPM in~\eqref{hpsm} can be approximated as
\begin{equation}\label{beamrep}
\mathbf{H}_{\text{HSPM}} \approx \overline{\mathbf{A}}_R \overline{\mathbf{H}}_b \overline{\mathbf{A}}_{T}^{\mathsf{H}},
\end{equation}
where \(\overline{\mathbf{H}}_b \in \mathbb{C}^{N_r \times N_t}\) is the sparse transformed matrix. While real channels may not be perfectly sparse, this representation still offers significant sparsity benefits~\cite{hpsm_yuhan}.
\begin{figure}[t]
    \centering
    \includegraphics[width=0.45\textwidth]{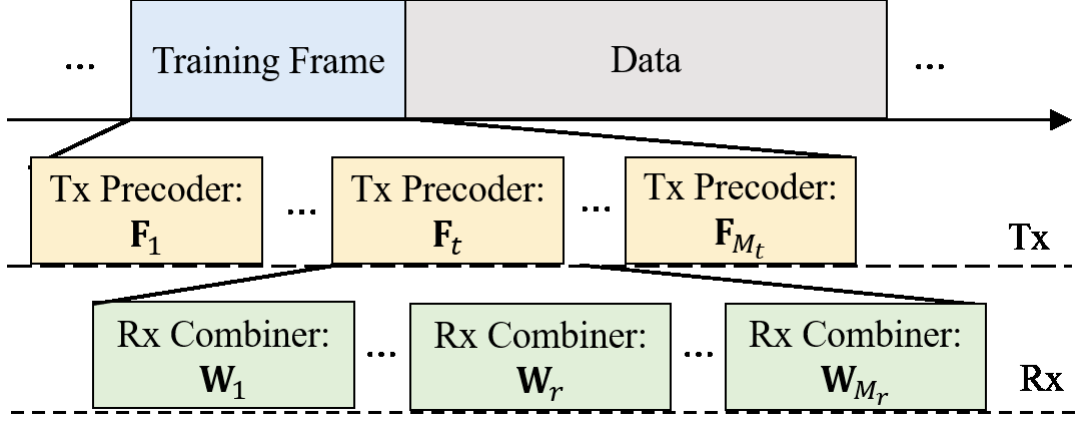}
    \caption{{Frame structure for the training process of CE.}}
    \label{fig_frame}
\end{figure}

\begin{figure*}[t]
    \centering
    \includegraphics[width=1.0\textwidth]{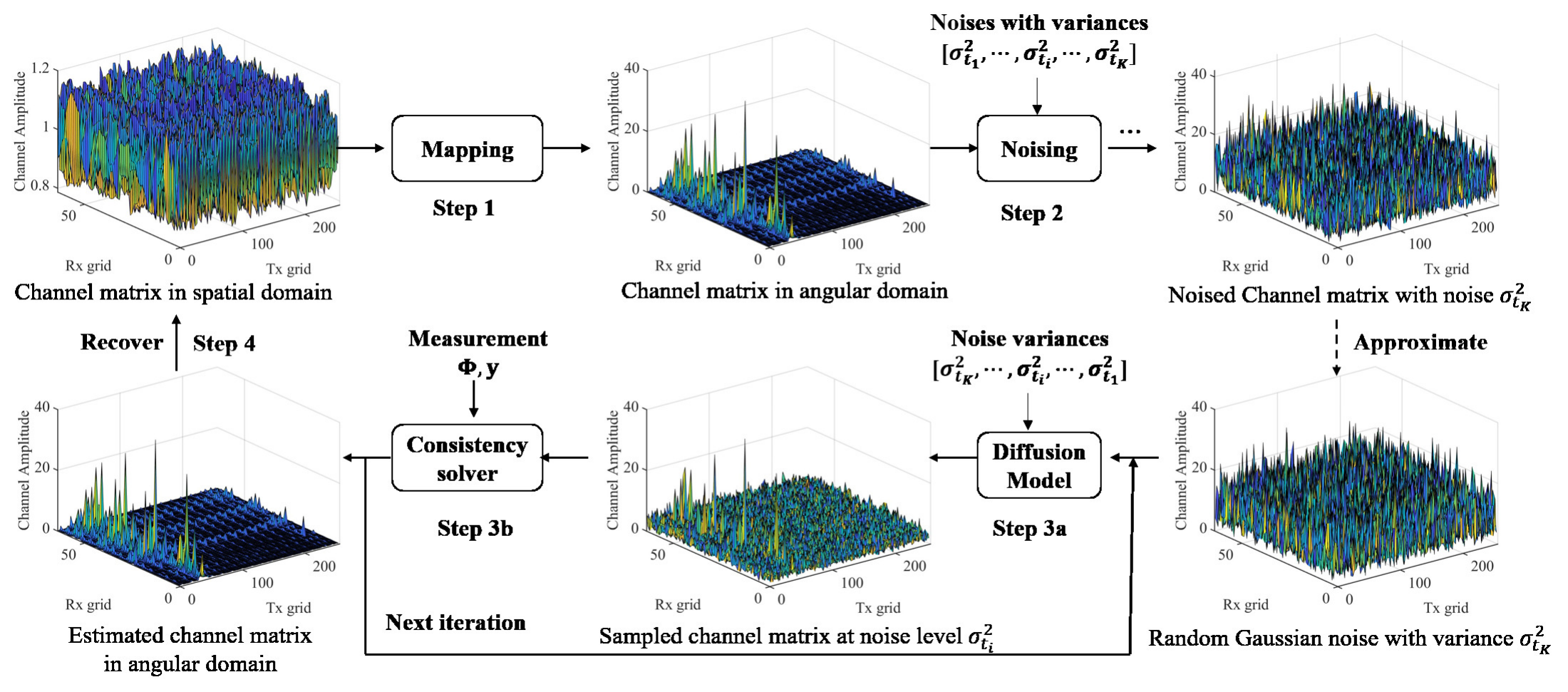} 
    \caption{Framework of diffusion-based plug-and-play augmented CE method.}
    \label{fig_method}
\end{figure*}

\subsection{CE Problem Formulation}
The frame structure adopted for transmission is shown in Fig.~\ref{fig_frame}. During the training process for CE, pilot signals are transmitted from the Tx in $M=M_tM_r$ training slots, where $M_t$ and $M_r$ denote the number of training slots allocated to Tx and Rx, respectively. For the \((r,t)^{\rm th}\) training slot, the received signal can be represented by
\begin{equation}\label{train_process}
\mathbf{y}_{r,t} ={\mathbf{W}}_{r}^{*} \mathbf{H} {\mathbf{F}}_{{t}} \mathbf{s}_{t} + {\mathbf{n}}_{r,t},
\end{equation}
where \(t = 1, \dots, M_t\) and \(r = 1, \dots, M_r\) index the training slot. The training combiner \(\mathbf{W}_{r} \in \mathbb{C}^{N_r \times L_r}\) and the training precoder \(\mathbf{F}_{t}\in \mathbb{C}^{N_t \times L_t}\) are used. The pilot symbol \(\mathbf{s}_{t} \in \mathbb{C}^{N_s \times 1}\) satisfies the power constraint \(\mathbb{E}\{\mathbf{s}_{t} \mathbf{s}_{t}^{*}\} = \frac{P}{N_s} \mathbf{I}_{N_s}\), with total power \(P\) and \(N_s = L_t\). The noise vector \(\mathbf{n}_{r,t}\in \mathbb{C}^{L_r \times 1}\) represents the combined noise in the \((r,t)^{\rm th}\) training slot.

Initially, training is conducted on the Rx side, and $M_r$ different combiners are employed to obtain received signal $\mathbf{y}_{t}$ as $\mathbf{y}_{t}=[\mathbf{y}_{1,t}^T,\cdots,\mathbf{y}_{M_r,t}^T]^T\in\mathbb{C}^{M_rL_r}$, which can be represented as
\begin{equation}
    \mathbf{y}_{t} = \mathbf{W}^{*} \mathbf{H} {\mathbf{F}}_{{t}} \mathbf{s}_{t} + {\mathbf{n}}_{t},
\end{equation}
where $ \mathbf{W}=[{\mathbf{W}}_1,\cdots,{\mathbf{W}}_{M_r}]\in\mathbb{C}^{N_r\times L_rM_r}$ denotes the training combiner, and $\mathbf{n}_{t}=[\overline{\mathbf{n}}_{1,t}^T,\cdots,\overline{\mathbf{n}}_{M_r,t}^T]^T\in\mathbb{C}^{M_rL_r}$.

After one round of Rx training, the Tx adjusts its training precoder $\mathbf{F}_t$. The transmitted pilot symbol can be represented as $\mathbf{p}_t=\mathbf{F}_t\mathbf{s}_{t}\in \mathbb{C}^{N_t\times 1}$. By collecting $\mathbf{y}_t$ for $t=1,\cdots,M_t$, the resulting received signals can be represented by $\mathbf{Y}=[\mathbf{y}_1,\cdots,\mathbf{y}_{M_t}]\in\mathbb{C}^{M_rL_r\times M_t}$. The resulting received signal is represented as
\begin{equation}
     \mathbf{Y} = \mathbf{W}^{*} \mathbf{H} \mathbf{P} + \mathbf{N},
\end{equation}
where $\mathbf{P}=[\mathbf{p}_1,\cdots,\mathbf{p}_{M_t}]\in \mathbb{C}^{N_t\times M_t}$ denotes the collection of transmitted pilot vectors, and $\mathbf{N}=[\mathbf{n}_1,\cdots,\mathbf{n}_{M_t}]\in \mathbb{C}^{M_rL_r\times M_t}$ denotes the received noise matrix. By inserting the beamspace representation in \eqref{beamrep}, the received signal can be represented as
\begin{equation}
     \mathbf{Y} = \mathbf{W}^{*} \overline{\mathbf{A}}_R \overline{\mathbf{H}}_b \overline{\mathbf{A}}_T^*\mathbf{P} + \mathbf{N}.
\end{equation}
Then, the received signal after vectorization can be written as
\begin{equation}
\mathbf{y} = \mathbf{\Phi}\mathbf{h}+ \mathbf{n},
\end{equation}
where \(\mathbf{y} = \mathrm{vec}(\mathbf{Y})\in\mathbb{C}^{M_tM_rL_r}\) represents the vectorized received signal, \(\mathbf{\Phi} = ((\overline{\mathbf{A}}_T^*\mathbf{P})^T \otimes \mathbf{W}^{*} \overline{\mathbf{A}}_R)\in\mathbb{C}^{M_tM_rL_r\times N_t N_r}\) denotes the measurement matrix, \(\mathbf{h} = \mathrm{vec}(\overline{\mathbf{H}}_b)\) denotes the vectorized beamspace channel, and \(\mathbf{n} = \mathrm{vec}(\mathbf{N})\) represents the vectorized noise. Then, the goal of the CE problem is to recover the sparse beamspace channel vector $\mathbf{h}$ from the received signal measurements $\mathbf{y}$. This can be formulated as a sparse signal recovery problem

\begin{equation}  
\min_{\mathbf{h}} \|\mathbf{h}\|_1 \quad \text{subject to} \quad \frac{1}{2\sigma_n^2}\| \mathbf{y} - \mathbf{\Phi}\mathbf{h}\|_2^2 \leq \epsilon, \label{eq:sparse_recovery}  
\end{equation}  
where $\|\mathbf{h}\|_1$ enforces sparsity via the $\ell_1$-norm, $\sigma_n^2$ is the noise variance, and $\epsilon$ controls the allowed estimation error. However, in practical scenarios, the sparsity assumption may not hold exactly due to grid mismatch and approximation errors in the HPSM model. As a result, relying solely on hand-crafted sparsity priors often fails to capture the true underlying structure of real-world channels. To address this limitation, we incorporate learned prior knowledge as an advanced regularization term, leading to the following optimization problem
\begin{equation}  
\hat{\mathbf{h}} = \underset{\mathbf{h}}{\arg\min} \, \frac{1}{2\sigma_n^2}\left\| \mathbf{y} - \mathbf{\Phi}\mathbf{h} \right\|_2^2 + \lambda \mathcal{P}(\mathbf{h}), \label{eq:problem}  
\end{equation}  
where $\hat{\mathbf{h}}$ is the estimated channel, $\frac{1}{2\sigma_n^2}\left\| \mathbf{y} - \mathbf{\Phi}\mathbf{h} \right\|_2^2 $ represents a consistency term, which ensures that the estimated channel $\mathbf{h}$ remains consistent with the observed measurements $\mathbf{y}$, $\mathcal{P}(\mathbf{h})$ represents a prior learned via a neural network, and $\lambda$ balances the trade-off between data fidelity and regularization.

\section{Diffusion-based Plug-and-play Augmented CE Method}\label{sec:method}
In this section, we propose a diffusion model-based plug-and-play augmented CE framework, as illustrated in Fig.~\ref{fig_method}.
\hdel{The overall approach addresses the optimization problem in~\eqref{eq:problem} through four successive steps. 
First, the spatial-domain channel matrix is transformed into a sparser angular-domain representation via the codebook projection in~\eqref{beamrep}, .
which reduces modeling complexity. 
Second, a diffusion model is employed to learn the underlying distribution of angular-domain channels, capturing real-world propagation characteristics that conventional sparsity priors cannot represent; the inference is performed by solving an ODE, enabling efficient sampling. 
Third, plug-and-play optimization alternates between enforcing measurement consistency through the data fidelity term $\tfrac{1}{2\sigma_n^2}\|\mathbf{y}-\mathbf{\Phi}\mathbf{h}\|_2^2$ and applying the learned diffusion prior via iterative denoising. 
Finally, the estimated channel is reconstructed in the spatial domain for practical use. 
The diffusion model architecture and consistency solver implementation are detailed in the following.}\revise{The overall approach addresses the optimization problem in~\eqref{eq:problem} through four successive steps. 
First, the spatial-domain channel matrix is transformed into a sparser angular-domain representation via the codebook projection in~\eqref{beamrep}, which reduces modeling complexity. 
Second, a diffusion model is employed to learn the underlying distribution of angular-domain channels, capturing real-world propagation characteristics that conventional sparsity priors cannot represent. 
Third, plug-and-play optimization alternates between enforcing measurement consistency through the data fidelity term $\frac{1}{2\sigma_n^2}\|\mathbf{y}-\mathbf{\Phi}\mathbf{h}\|_2^2$ and applying the learned diffusion prior via iterative denoising. 
Finally, the estimated channel is reconstructed in the spatial domain for practical use. 
The diffusion model architecture and consistency solver implementation are detailed in the following.
}

\subsection{Diffusion Denoiser}\label{sec:diffusion_design}
The diffusion model is one of the most robust generative models, using a reverse diffusion process to iteratively transform a random noise distribution into a target data distribution. It can also be interpreted as a sophisticated denoiser, which eliminates noise that lies outside the desired channel distribution. Through this step-by-step refinement, the final denoised samples align closely with the intended channel distribution, achieving high-quality generative results. The detailed training process of the diffusion model is introduced as follows. 

Given a channel distribution $\mathcal{P}(\mathbf{h})$, the distribution is perturbed by progressively adding noise with increasing variances \( \sigma_{t_1}^2 < \sigma_{t_2}^2 < \cdots < \sigma_{t_K}^2 \). The variable \( t_i \in \{t_1,\dots,t_K\} \) denotes discrete time points sampled from
the continuous interval \( [0,T] \). This process of adding noise is called a diffusion process, which gradually drifts the target distribution towards a random Gaussian distribution. The mathematical representation of this process can be formulated as
\begin{equation}
    \mathbf{h}_{t_i} = \mathbf{h}_{t_{i-1}}+\sqrt{\sigma_{t_i}^2-\sigma_{t_{i-1}}^2} \mathbf{n}_{i-1},
\end{equation}
where $\mathbf{h}_{t_i}$ represents the perturbed sample with noise variance $\sigma_{t_i}^2$, and $\mathbf{n}_{i-1}\sim\mathcal{CN}(0,\mathbf{I}_{N_tN_r})$ represents the added noise to samples. As $K \to \infty$, the sequence $\{\sigma_{t_i}\}$ converges to a continuous function $\sigma(t)$ with the continuous time variable $t \in [0,T]$. Specifically, the linear parameterization 
$\sigma(t) = t$ is employed throughout this work, as empirically it demonstrates good performance according to~\cite{edm}. As a result, the sequence $\{\mathbf{h}_{t_i}\}$ becomes a stochastic process $\{\mathbf{h}(t)\}_{t=0}^T$. For notational convenience, the variables $\mathbf{h}(t)$ and $\sigma(t)$ are denoted by $\mathbf{h}_t$ and $\sigma_t$, respectively.
 The stochastic process can then be represented by
\begin{equation}
    \mathrm{d\mathbf{h}}_t = \sqrt{\frac{\mathrm{d}\sigma^2_t}{\mathrm{d}t}}\mathrm{d}\mathbf{w},
\end{equation}
where $\mathbf{w}\sim \mathcal{CN}(0,t)$ denotes the complex-valued Wiener process, and $\mathrm{d}\mathbf{w}\sim\mathcal{CN}(0,dt)$ is the differential of the Wiener process. As the noise variance increases, the noised channel matrix approaches a random Gaussian noise.

Then, the training target of the diffusion denoiser is to recover the channel distribution from the random noise distribution. This process is called a reverse process, which moves the perturbed samples to the target channel distribution. Through the denoising process, the knowledge of the channel distribution can be learned by the denoiser, which can serve as prior knowledge for the CE problem. The stochastic process can be approximately reversed by a deterministic process, which can be characterized by an ODE. Notably, this deterministic sampling approach by solving ODE can significantly accelerate sampling speed of diffusion models compared to stochastic sampling methods~\cite{edm}. The denoised sample can be calculated by the following ODE
\begin{equation}\label{ode}
    \mathrm{d}\mathbf{h}_t =-\dot{\sigma}_t\sigma_t\nabla_{\mathbf{h}_t}\mathrm{log} p(\mathbf{h}_t;\sigma_t)\mathrm{d}t,
\end{equation}
where $\dot{\sigma}_t$ represents the derivative of $\sigma_t$ with respect to the time $t$, and $\nabla_{\mathbf{h}_t}\mathrm{log}p(\mathbf{h}_t;\sigma_t)$ denotes the gradient of the log-probability density with respect to $\mathbf{h}_t$. The latter term, known as the score function of the marginal distribution, points in the direction of steepest increase in the probability density. The Euler method can be used to solve the ODE, which can be expressed as
\begin{equation}\label{solver}
    \mathbf{h}_{t-\Delta t}=\mathbf{h}_t+\Delta t\cdot\dot{\sigma}_t\sigma_t\nabla_{\mathbf{h}_t}\mathrm{log} p(\mathbf{h}_t;\sigma_t),
\end{equation}
where $\Delta t$ represents the step size in solving the ODE.

To instantiate the ODE solver for reversing the stochastic process, the score function must first estimated through training. The training objective can be defined as
\begin{align}
\theta^* = \arg\min_{\theta} \mathbb{E}_{\mathbf{h}_t,\sigma_t}\left[ \left\| \mathbf{s}_{\theta}(\mathbf{h}_t, \sigma_t) - \nabla_{\mathbf{h}_t} \log p(\mathbf{h}_t; \sigma_t) \right\|_2^2 \right],
\end{align}
where $\mathbf{s}_\theta$ denotes the diffusion network with parameters $\theta$. The network takes perturbed samples $\mathbf{h}_t$ and corresponding noise level $\sigma_t$ as input. Minimizing this loss encourages the network output to approximate the score function $\nabla_{\mathbf{h}_t} \log p(\mathbf{h}_t;\sigma(t))$. Since $\mathbf{h}_t$ is obtained by adding Gaussian noise with variance $\sigma_t^2$ to $\mathbf{h}$, the score function has a closed-form expression as
\begin{equation}
\nabla_{\mathbf{h}_t} \log p(\mathbf{h}_t;\sigma_t) = -\frac{\mathbf{h}_t - \mathbf{h}}{\sigma_t^2}.
\end{equation}
Substituting this into the loss function yields
\begin{equation}
\theta^* = \arg\min_{\theta} \mathbb{E}_{\mathbf{h}_t,\sigma_t}\left[ \left\| \mathbf{s}_{\theta}(\mathbf{h}_t, \sigma_t) + \frac{\mathbf{h}_t - \mathbf{h}}{\sigma_t^2} \right\|_2^2 \right].
\end{equation}
By introducing a residual connection that adds the input directly to the output, the diffusion network admits an equivalent parameterization as a denoiser network $\mathbf{D}_\theta$ defined by
\begin{equation}
\mathbf{D}_{\theta}(\mathbf{h}_t, \sigma_t) = \mathbf{h}_t + \sigma_t^2 \mathbf{s}_{\theta}(\mathbf{h}_t, \sigma_t).
\end{equation}
This transforms the loss into a denoising objective, resulting in
\begin{equation}\label{eq_loss}
\theta^* = \arg\min_{\theta} \mathbb{E}_{\mathbf{h}_t,\sigma_t}\left[ \left\| \mathbf{D}_{\theta}(\mathbf{h}_t, \sigma_t) - \mathbf{h} \right\|_2^2 \right].
\end{equation}
After training, the score function can be directly derived from the denoiser network as
\begin{equation}\label{eq_denoiser}
\mathbf{s}_{\theta}(\mathbf{h}_t, \sigma_t) = \frac{\mathbf{D}_{\theta}(\mathbf{h}_t, \sigma_t) - \mathbf{h}}{\sigma_t^2}.
\end{equation}
This score function provides an essential prior for the CE problem, as it enables the reconstruction of the channel distribution $\mathcal{P}(\mathbf{h})$ from Gaussian noise through the reverse diffusion process governed by the ODE in~\eqref{ode}.

\begin{figure}[t]
    \centering
\includegraphics[width=0.5\textwidth]{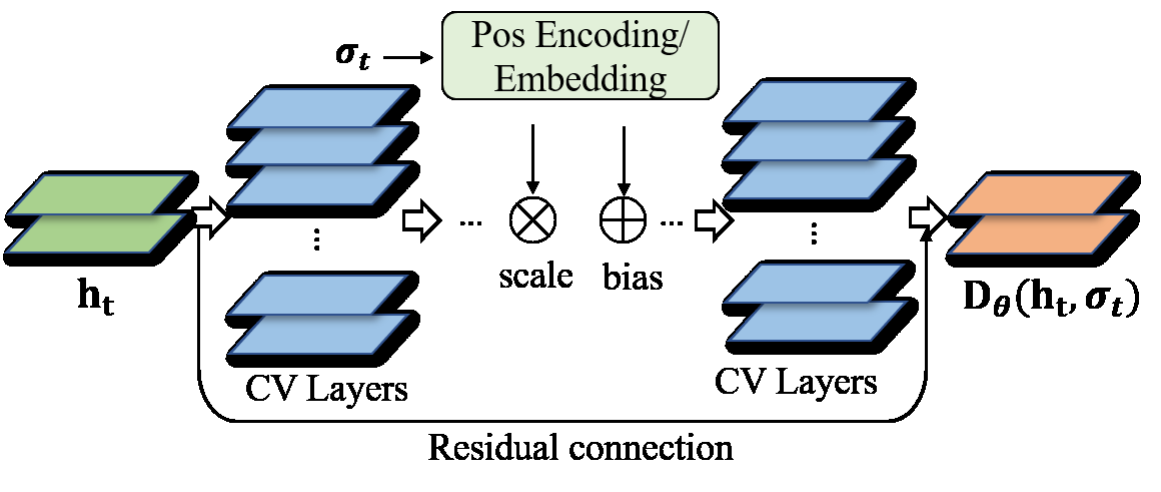} 
    \caption{Network structure of the diffusion denoiser.}
    \label{network}
\end{figure}

\subsection{Diffusion Network Design}

In mmWave and THz UL-MIMO systems, the high dimensionality of channel matrices necessitates an efficient diffusion model architecture for practical deployment. We employ a lightweight CNN architecture~\cite{diffusion_cnn} that removes computationally intensive components, such as attention mechanisms. While these mechanisms can enhance feature extraction in frameworks like SongUNet~\cite{songunet}, they incur prohibitive memory and latency costs for real-time wireless systems.

As shown in Fig.~\ref{network}, our network processes a 2D noised channel matrix (containing real and imaginary components) alongside the noise level \( \sigma_t \), a critical parameter guiding denoising. The noise level first undergoes positional encoding to capture its temporal properties and then is transformed through linear layers into an embedded noise representation. This embedding is used to adaptively scale and shift the feature maps produced by each convolutional (CV) layer in the encoder. The decoder, also based on CV layers, reconstructs the denoised channel from these enhanced features. Both the encoder and decoder consist of four CV layers.

A residual connection adds the decoder's output to the original input, ensuring stable gradient flow during backpropagation while preserving critical low-level features for accurate reconstruction. This balanced architecture maintains computational efficiency while delivering high CE accuracy for THz and mmWave MIMO systems.

\subsection{Consistency Solver}\label{sec:solver}
Through the diffusion denoiser, channel samples can be generated that follow the distribution of the target channel. However, in solving the CE problem, it is not only important to ensure that the estimated channel lies within the target distribution, but it is also crucial to ensure that the estimated channel sample $\mathbf{h}$ is consistent with the received observation $\mathbf{y}$. Therefore, a consistency solver is necessary to guide the estimated channel towards the true channel behind the measurement observation.

To design the consistency solver, a plug-and-play approach is employed. This method leverages a pre-trained deep neural network as a denoiser, which is seamlessly integrated into an iterative optimization process that progressively converges toward the optimized solution. The term ``plug-and-play" highlights the framework's modular design, which enables seamless interchangeability of denoising operators while preserving the overall architecture. Specifically, the optimization process decouples the consistency term and the prior term in ~\eqref{eq:problem} into two subproblems, by introducing an auxiliary term $\mathbf{z}$,
\begin{algorithm}[t]
\caption{DiffPace}
\begin{algorithmic}[1]
\REQUIRE $\mathbf{s}_{\theta}, \mathbf{y}, K, \sigma_n, \{\sigma_{t_i}\}_{i=1}^K,\lambda,\beta$
\STATE Initialize $\mathbf{h}_{t_K} \sim \mathcal{CN}(0, \sigma_{t_K}^2\mathbf{I}_{N_rN_t})$
\FOR{$i = K$ \textbf{to} 1}
     \STATE $\Delta \sigma_{t_i} = \sigma_{t_i}({t_i}-{t_{i-1}})$
      \STATE $\rho_i =\Delta \sigma_{t_i}/({2\lambda\sigma_n^2+\beta\sigma_{t_i}^2})$
    \STATE $\mathbf{z}_{i} = \mathbf{h}_{t_i} + \Delta \sigma_{t_i}\cdot\mathbf{s}_{\theta}(\mathbf{h}_{t_i}, \sigma_{t_i})$
    \STATE $\mathbf{h}_{t_{i-1}}=\mathbf{z}_{i}+\rho_i\mathbf{\Phi}^T(\mathbf{\Phi}\mathbf{\Phi}^T)^{-1}(\mathbf{y}-\mathbf{\Phi} \mathbf{z}_i)$
\ENDFOR
\RETURN $\hat{\mathbf{h}}=\mathbf{h}_{t_0}$
\end{algorithmic}
\end{algorithm}
\begin{subequations}
 \begin{empheq}[left=\empheqlbrace]{align}
    \mathbf{z}_i &= \underset{\mathbf{z}}{\arg \min}~\mu\| \mathbf{z} - \mathbf{h}_{t_i} \|^2 + \lambda\mathcal{P}(\mathbf{z}), \label{eq_sub1} \\ 
    \mathbf{h}_{t_{i-1}} &= \underset{\mathbf{h}}{\arg \min} \frac{1}{2\sigma_n^2}\| \mathbf{y} - \mathbf{\Phi} \mathbf{h} \|^2 + \mu  \| \mathbf{h} - \mathbf{z}_i \|^2, \label{eq_sub2}
\end{empheq}
\end{subequations}
where \( i = 1, \dots, K \) enumerates the optimization iterations, and \( \mu\) is a regularization 
parameter that enforces data consistency between the subproblems. Equation \eqref{eq_sub1} represents the prior term, which ensures that the estimated samples \( \mathbf{z}_i \) remain close to the desired distribution. This subproblem can be solved by the diffusion denoiser, which removes the noise in $\mathbf{h}_{t_i}$ and recovers the sample within the channel distribution. The solution to this subproblem is given by
\begin{equation}
    \mathbf{z}_i = \mathbf{h}_{t_i} +\Delta \sigma_{t_i}\cdot\mathbf{s}_{\theta}(\mathbf{h}_{t_i},\sigma_{t_i}),
\end{equation}
where $\mathbf{s}_{\theta}(\mathbf{h}_{t_i},\sigma_{t_i})=\nabla_{\mathbf{h}_{t_{i}}} p(\mathbf{h}_{t_i};\sigma_{t_i})$ represents the score function estimated by the diffusion network, which pulls the estimated sample toward the high-density region of the channel distribution at noise scale $\sigma_{t_i}$. The term $\Delta \sigma_{t_i}=\lambda/{\mu}$ corresponds to the discrete Euler step $\Delta t\cdot\dot{\sigma}_t\sigma_t $ from equation~\eqref{solver}. For the linear noise schedule $\sigma_{t_i} = t_i$ with $\Delta t_i = t_{i} - t_{i-1}$, this step length satisfies
\begin{equation}
    \Delta \sigma_{t_i}=\Delta t_i\cdot\dot{\sigma}_{t_i}\sigma_{t_{i}}=\sigma_{t_i}({t_i}-{t_{i-1}}).
\end{equation}
From the original definition $\Delta \sigma_{t_i} = \lambda/\mu$, the term $\mu$ is given by 
$\mu=\lambda/{\Delta \sigma_{t_i}}$.

Meanwhile, \eqref{eq_sub2} enforces the consistency term by minimizing \( \| \mathbf{y} - \mathbf{\Phi h} \|^2 \). This problem can be solved by projecting $\mathbf{z}_t$ onto the subspace defined by the constraint $\mathbf{y}=\mathbf{\Phi} \mathbf{h}$ as
\begin{equation}
    \mathbf{h}_{t_{i-1}}=\mathbf{z}_i +\rho_i \mathbf{\Phi}^T(\mathbf{\Phi}\mathbf{\Phi}^T)(\mathbf{y}-\mathbf{\Phi} \mathbf{z}_t).
\end{equation}
This formula adjusts \(\mathbf{z}_i\) by a correction term that aligns it with the constraint \(\mathbf{y}=\mathbf{\Phi}\mathbf{h}\), ensuring minimal deviation from \(\mathbf{z}_i\). The step length of the correction term is given by $\rho_i ={1}/({2\sigma_n^2\mu})={\Delta \sigma_{t_i}}/({2\lambda\sigma_n^2})$. In practice, we regularize the expression of $\rho_i$ by adding a smoothing term $\beta\sigma_t^2$ to the denominator to prevent numerical instability, yielding $\rho_i ={\Delta \sigma_{t_i}}/({2\lambda\sigma_n^2+\beta\sigma_{t_i}^2})$. \revise{The parameters 
$\lambda$ and $\beta$ are chosen empirically via grid search on a validation set. Intuitively, $\lambda$ controls the overall strength of the correction step. A smaller $\lambda$ increases the step size $\rho_i$, accelerating alignment with the constraint but potentially causing instability, while a larger $\lambda$ leads to more conservative updates. The parameter $\beta$ stabilizes the denominator when $\sigma_t$ becomes small, preventing excessively large correction terms and improving robustness in the high-SNR regime.} By tuning the parameters \(\lambda\) and \(\beta\), the step size \(\rho_i\) adapts to both the step lengths of the ODE solver and the noise variance of the channel.

By iteratively solving these two sub-problems, the channel $\mathbf{h}$ can be estimated. Specifically, the iterative rule can be represented by
\begin{subequations}
 \begin{empheq}[left=\empheqlbrace]{align}
    \mathbf{z}_i &=   \mathbf{h}_{t_i} +\Delta \sigma_{t_i}\cdot\mathbf{s}_{\theta}(\mathbf{h}_{t_i},\sigma_{t_i}),\label{eq_update1} \\ 
    \mathbf{h}_{t_{i-1}} &= \mathbf{z}_t +\rho_i\mathbf{\Phi}^T(\mathbf{\Phi}\mathbf{\Phi}^T)(\mathbf{y}-\mathbf{\Phi} \mathbf{z}_i).\label{eq_update2}
\end{empheq}
\end{subequations}
By incorporating the updating rule, the procedure for CE is provided in $\textbf{Algorithm 1}$.

\section{Results and Performance}\label{sec:evaluation}
In this section, we provide a detailed description of the simulation setup and thoroughly explain the training procedure for the diffusion model. The performance of the proposed method is then evaluated in terms of convergence, estimation accuracy,
and computational complexity. The accuracy of channel estimation is characterized by the normalized mean squared error (NMSE), which is given by 
\newcommand{\norm}[1]{\left\lVert#1\right\rVert}
\begin{equation}
    {\rm NMSE} = \frac{\norm{\mathbf{h}-\hat{\mathbf{h}}}_F^2}{\norm{\mathbf{h}}_F^2}.
\end{equation}

\begin{figure*}[t]
		\centering
		\subfigure[60~GHz channel from the Raymobtime datasets~\cite{b23}.
		]{\includegraphics[width=0.45\textwidth,height=4cm]{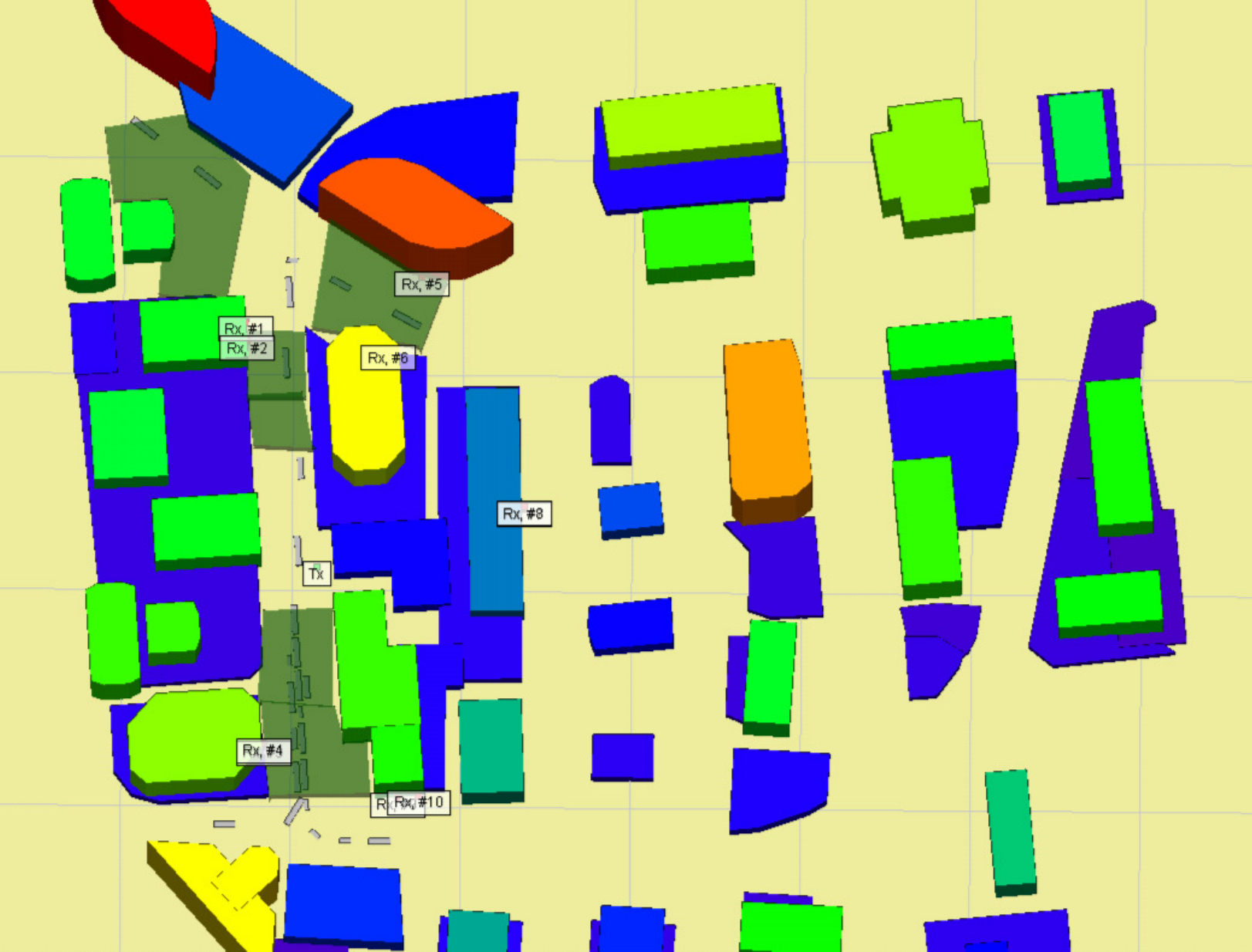}}
		\subfigure[0.3 THz channel from QuaDRiGa with measurement layout in the indoor corridor scenario~\cite{yuanbo_icc}.]{\includegraphics[width=0.45\textwidth,height=4cm]{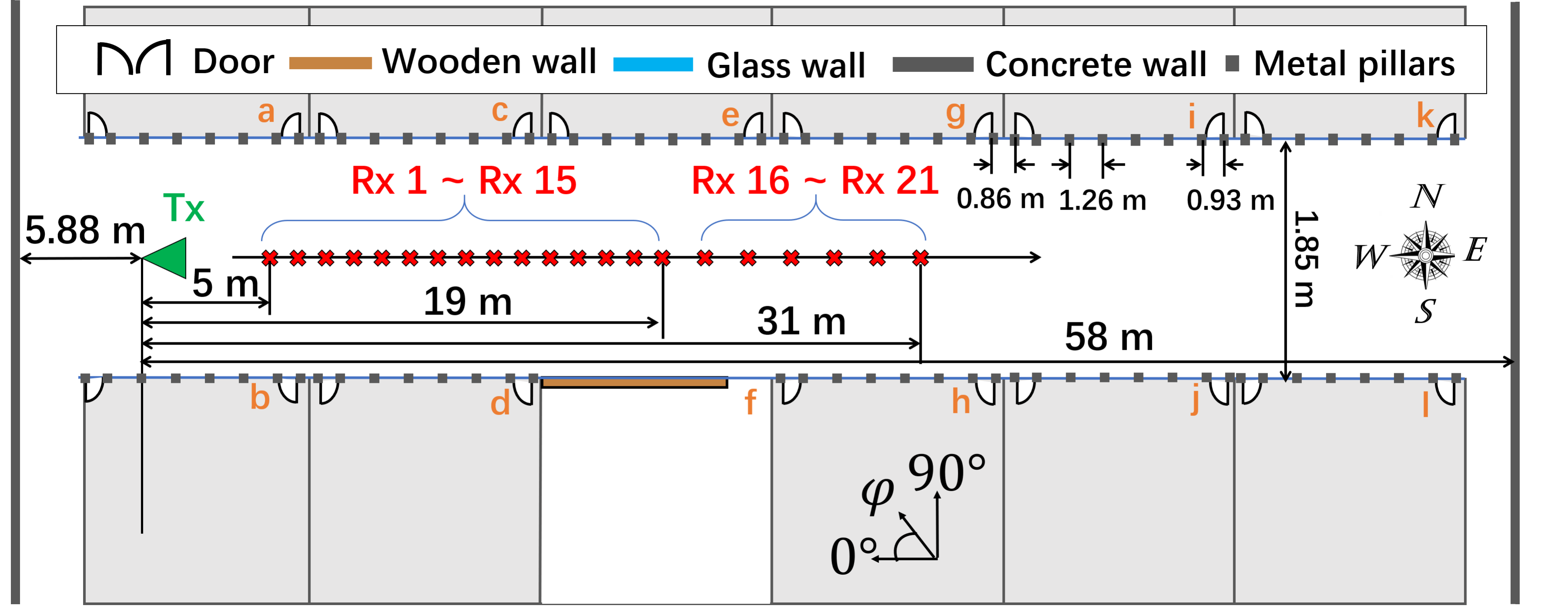}} 
		\caption{60 GHz and 0.3 THz channel propagation environment for Raymobtime dataset and Quadriga simulator.}
		\label{fig_simulation} 
\end{figure*}
\subsection{Dataset and Simulation Setup}
The proposed DiffPace framework is evaluated using both mmWave (60~GHz) and THz (0.3~THz) channel datasets. The mmWave channel is derived from the Raymobtime dataset~\cite{b23}, generated through 3D ray-tracing simulations that capture receiver mobility and temporal dynamics. Our analysis focuses on the s002 subset which models propagation in Rosslyn with 10 fixed receiver positions, as shown in Figure~\ref{fig_simulation}a. From this subset, we extract 10,000 channel realizations for training and testing purposes. For the THz channel, we generate 2,000 channel realizations using QuaDRiGa~\cite{quadriga}, calibrated with real measurement statistics from indoor corridor scenarios at 306–321 GHz~\cite{yuanbo_icc}, as obtained by our group. \revise{Specifically, the calibration is performed by incorporating key channel statistics obtained from measurements, including the path loss exponent, the mean and standard deviation of the $K$-factor, the delay spread, the angular spread, and the spatial correlation matrix. 
These parameters are used by the geometry-based stochastic channel model (GSCM) within QuaDRiGa to synthesize channel multipath components that closely match the observed THz propagation environment.}\hdel{The channel model follows the 3GPP standardized channel model, while} The experimental corridor setup is shown in Figure~\ref{fig_simulation}b.

The channel matrices are constructed using HPSM by incorporating multipath components including the delays, angles, and powers from the channel datasets. This unified approach handles intra-subarray channels with planar wavefronts while modeling inter-subarray propagation through spherical waves, effectively eliminating the need for explicit near-field/far-field classification and enabling seamless operation across both regimes.

The simulation parameters are configured as follows. For the mmWave channel, the number of antennas, RF chains, and subarrays at the Tx and Rx are set to \(N_t = 64\), \(L_t = 2\), $K_t=2$, \(N_r = 16\), \(L_r = 4\), and $K_r=2$ respectively. For the THz channel, the corresponding parameters are \(N_t = 256\), \(L_t = 4\), $K_t=2$, \(N_r = 64\), \(L_r = 8\) and $K_r=2$. The phase shifter resolution is set to \(N_b = 4\) bits. To construct the measurement matrix \(\mathbf{\Phi}\), the symbol vector \(\mathbf{s}_t\) is generated by randomly selecting quadrature phase shift keying (QPSK) symbols. The cascaded precoder \(\mathbf{F}_{t}\) and combiner \(\mathbf{W}_{r}\) in (11) are constructed using random phase shifts, with each matrix entry having the same amplitude to satisfy the constant modulus constraint. During pilot transmission, the entries of the analog precoder and combiner are chosen as random phase shifts with constant modulus, providing a simple and hardware-friendly baseline widely used in practice. \revise{While this work adopts random beamformers, the plug-and-play design of DiffPace can naturally incorporate optimized pilot and beamformer schemes, which may further reduce overhead and improve estimation accuracy.} To account for the finite resolution of phase shifters, the phase values are selected from the discrete set \(\Omega = \{1, e^{j2\pi/2^{N_b}}, e^{j4\pi/2^{N_b}}, \cdots, e^{j2\pi(2^{N_b}-1)/2^{N_b}} \}\). \revise{In our experiments, $N_b = 4$ corresponds to 4-bit phase shifters. Using lower-resolution phase shifters (e.g., 2--3 bits) would further degrade the measurement quality. Nevertheless, the learned channel prior in DiffPace compensates for the reduced information, providing robustness to such hardware constraints and outperforming conventional estimators under these conditions.}

The datasets are divided into 90\% for training and 10\% for testing. After dataset generation, the diffusion network is trained using the adaptive moment estimation optimizer (Adam), chosen for its rapid convergence rate, with a learning rate of \(l_r = 10^{-4}\). The chosen values of $\lambda$ and $\beta$ are 0.006 and 0.05 for the mmWave channel, and 0.001 and 0.03 for the THz channel, respectively.

\begin{figure*}[t]
		\centering
		\subfigure[Training loss and testing loss versus number of epochs.
		]{\includegraphics[width=0.4\textwidth]{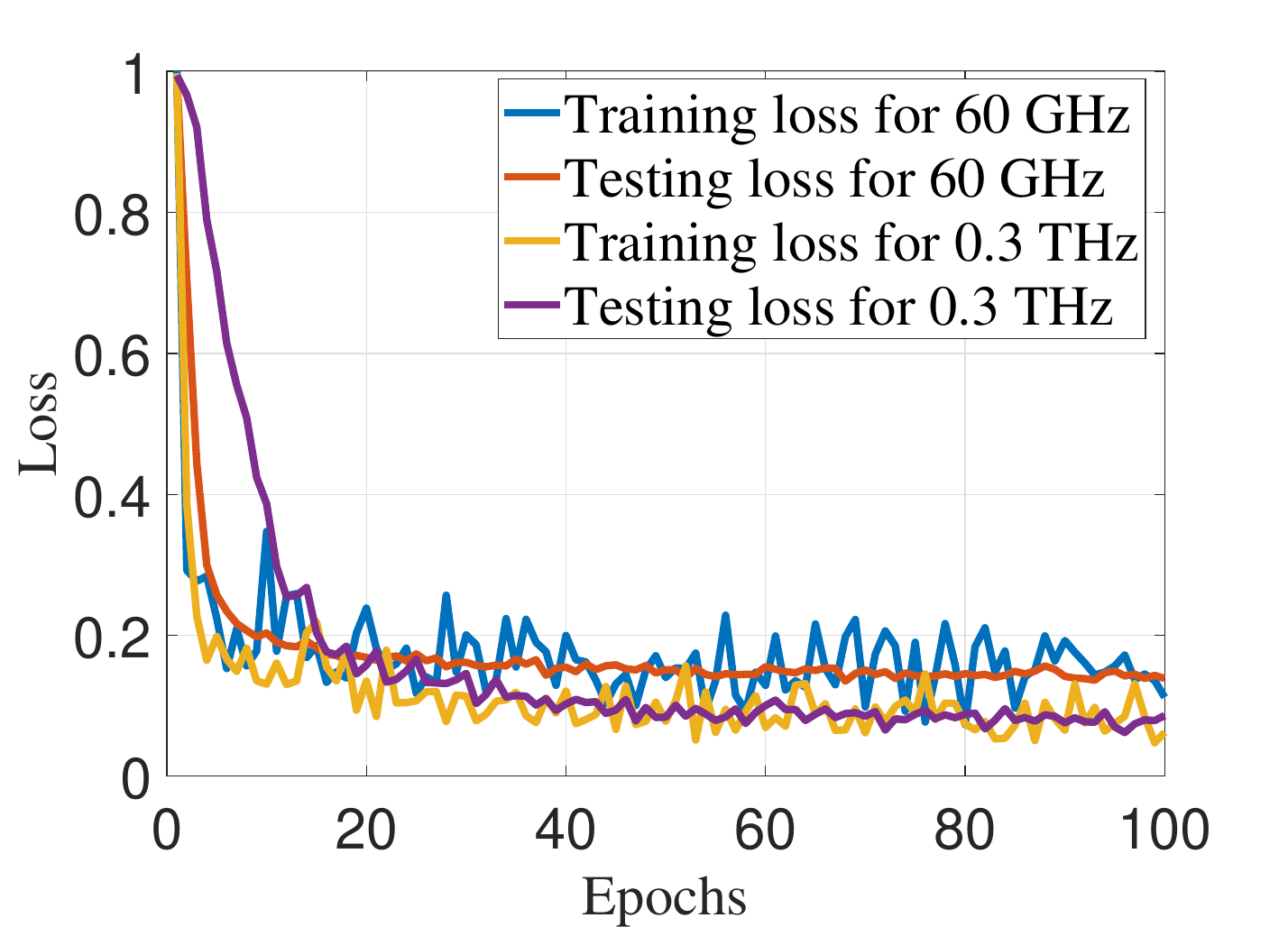}}
		\subfigure[Testing loss versus the proportion of utilized training samples.]{\includegraphics[width=0.4\textwidth]{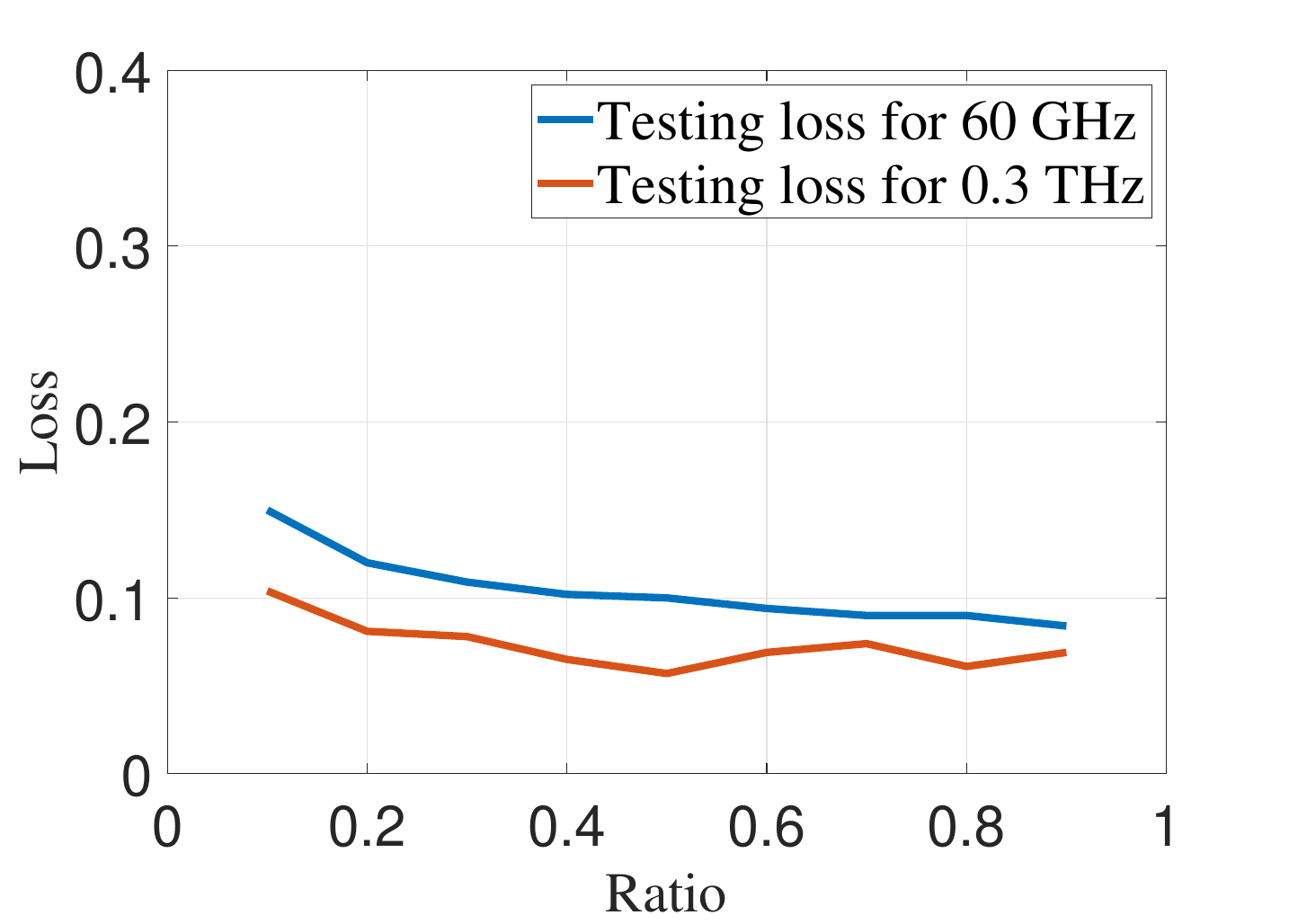}} 
		\caption{Convergence evaluation of DiffPace method for 60 GHz and 0.3 THz channel.}
		\label{fig_loss} 
\end{figure*}

\subsection{Training and Inference of Diffusion}
The diffusion model training consists of two fundamental processes, the forward noising process and the reverse denoising process. The noising process employs the noise schedule from the iDDPM method~\cite{iddpm}, implementing 1000 discrete noise levels uniformly applied to training samples. The model learns to denoise these corrupted samples through optimization of the loss function in \eqref{eq_loss}. To enhance training stability, we implement an exponential moving average (EMA) strategy for parameter updates, where the test-time network parameters $\theta'_j$ evolve according to $\theta'_j = m \theta'_{j-1} + (1-m)\theta_j$, with $\theta_j$ representing the parameters after the $j^{\mathrm{th}}$ training iteration and $m$ controlling the update rate. The stabilized parameters $\theta'_i$ address two objectives simultaneously, namely enabling model evaluation and serving as initialization for future training stages.

During inference, the ODE reverse process enables precision-complexity trade-offs through two key parameters, the denoising step count $K$ and the pilot ratio $\alpha = {M_t M_r L_r}/({N_t N_r})$. The pilot ratio quantifies the underdetermined nature of the measurement matrix $\mathbf{\Phi} \in \mathbb{C}^{M_t M_r L_r \times N_t N_r}$. In our experiments, we use $K = 100$ steps with $\alpha = 0.8$ to maintain a balance between computational load and estimation quality. The regularization coefficients $\lambda$ and $\beta$, governing prior weighting and step size respectively, undergo empirical optimization to guarantee convergence stability while preserving estimation accuracy.

All experiments have been executed on a workstation equipped with an AMD Ryzen Threadripper 3990X 64-Core Processor and four NVIDIA GeForce RTX 3090 GPUs, ensuring efficient processing of the computationally intensive diffusion operations.
\revise{In terms of training time, the diffusion model requires approximately 0.58 hours on the mmWave dataset with 10000 samples and 2.27 hours on the THz dataset with 2000 samples when trained for 100 epochs. This training is an offline process performed only once, after which the trained model can be reused across different pilot configurations and SNR conditions without the need for retraining.} \revise{Moreover, inference is highly efficient. Each denoising step takes about 0.001 seconds on a GPU and results in negligible latency. On a CPU, inference requires about 0.1 seconds per channel realization, which remains acceptable for non–real-time tasks such as offline channel sounding. For real-time deployment, lightweight GPUs or edge accelerators are preferable, and additional efficiency can be achieved through model compression or pruning techniques.} \revise{Furthermore, the growing adoption of GPU-based infrastructures in radio access networks, such as NVIDIA's Aerial AI framework, indicates that relying on GPU acceleration is both practical and promising for future wireless systems. These infrastructures provide a unified platform that supports AI-driven channel estimation together with traditional RAN functions~\cite{ai-ran}.}

\begin{figure*}[t]
    \centering
    \subfigure[60 GHz channel.]{\includegraphics[width=0.4\textwidth]{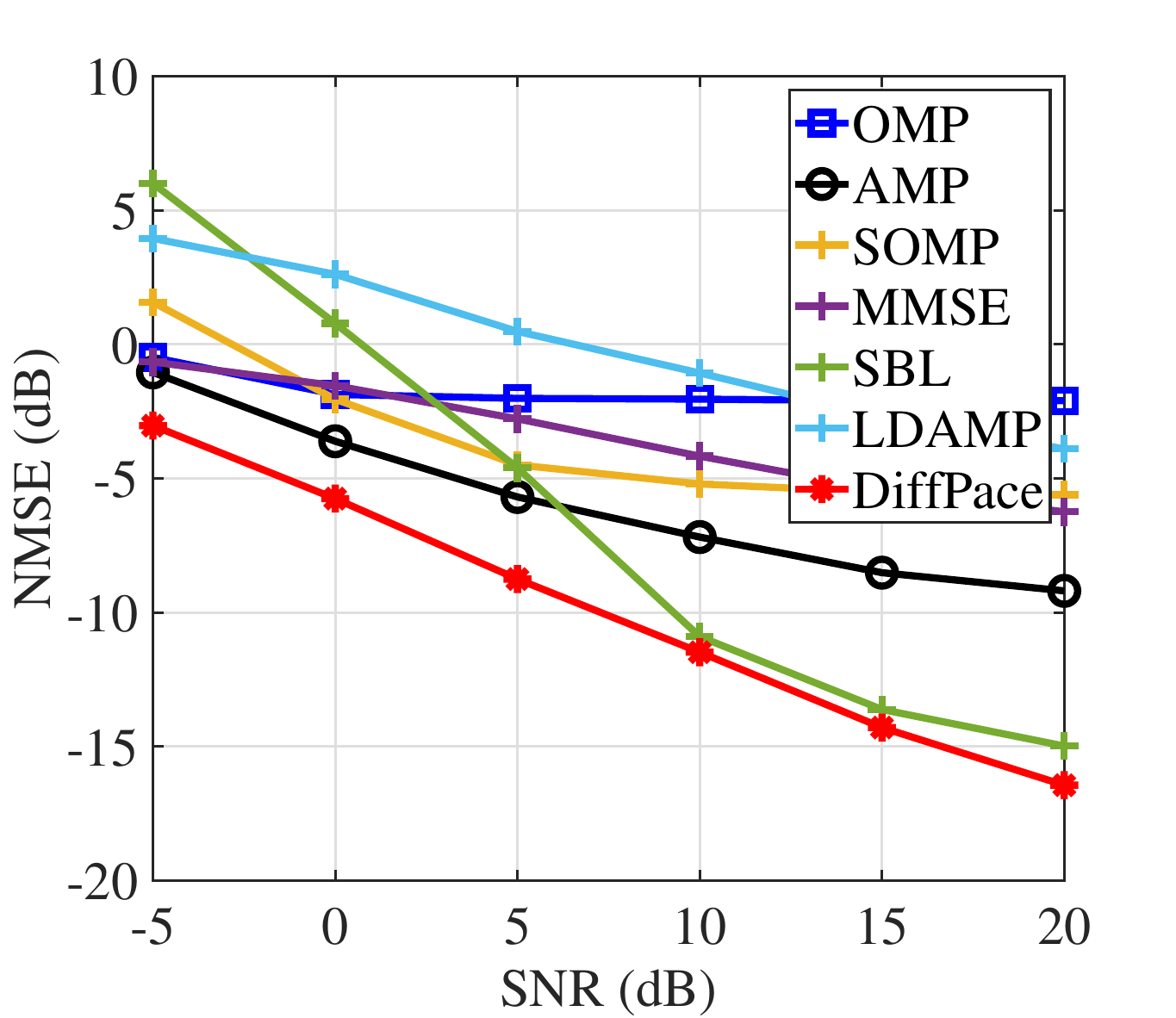}} 
    \subfigure[0.3 THz channel.]{ \includegraphics[width = 0.4\textwidth]{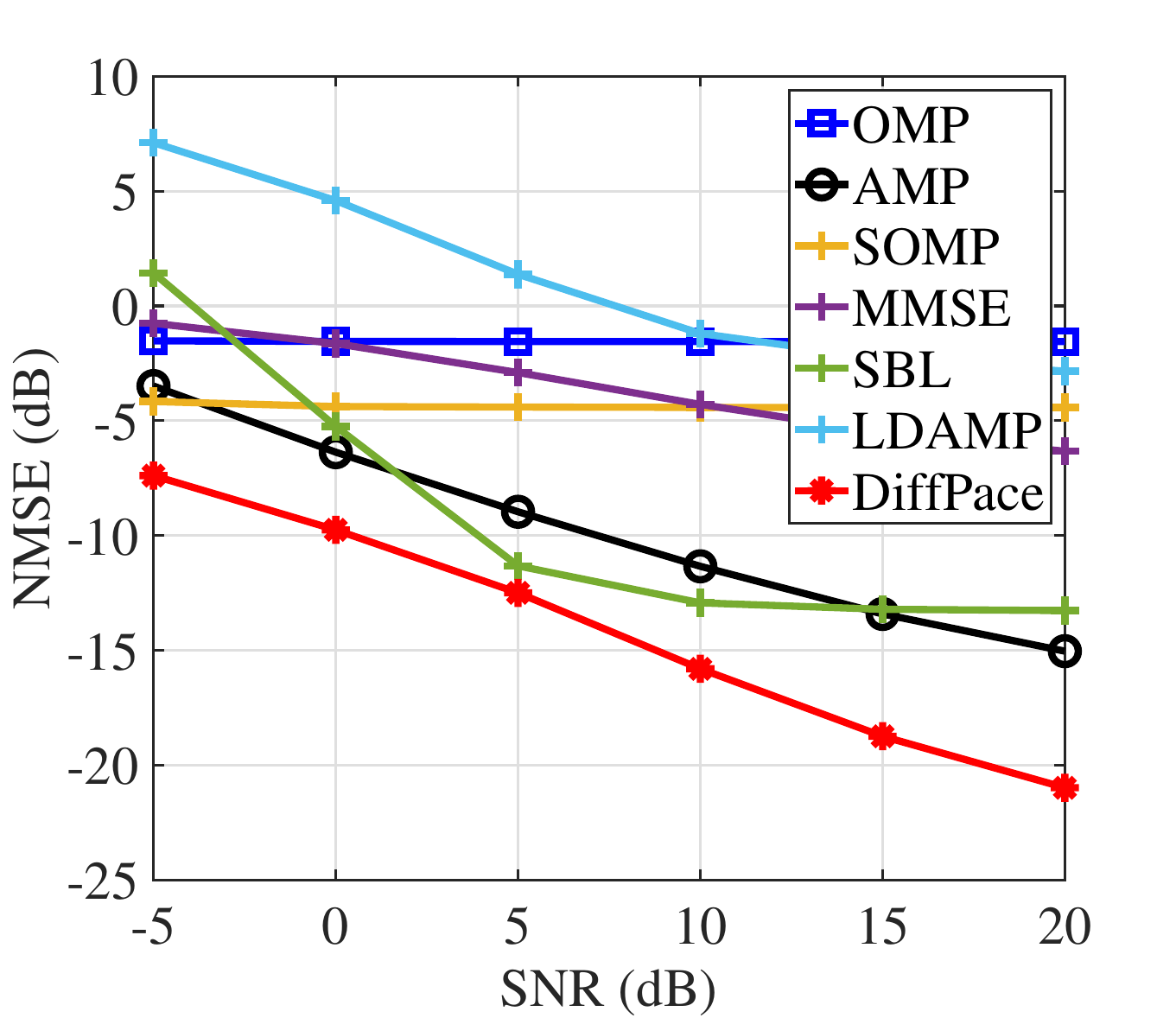}} 
    \caption{{CE accuracy of the DiffPace algorithm with $K=100$ and $\alpha=0.8$.}}
    \label{fig_accuracy}
\end{figure*}
\begin{figure*}[t]
    \centering
    \subfigure[60 GHz channel.]{\includegraphics[width=0.4\textwidth]{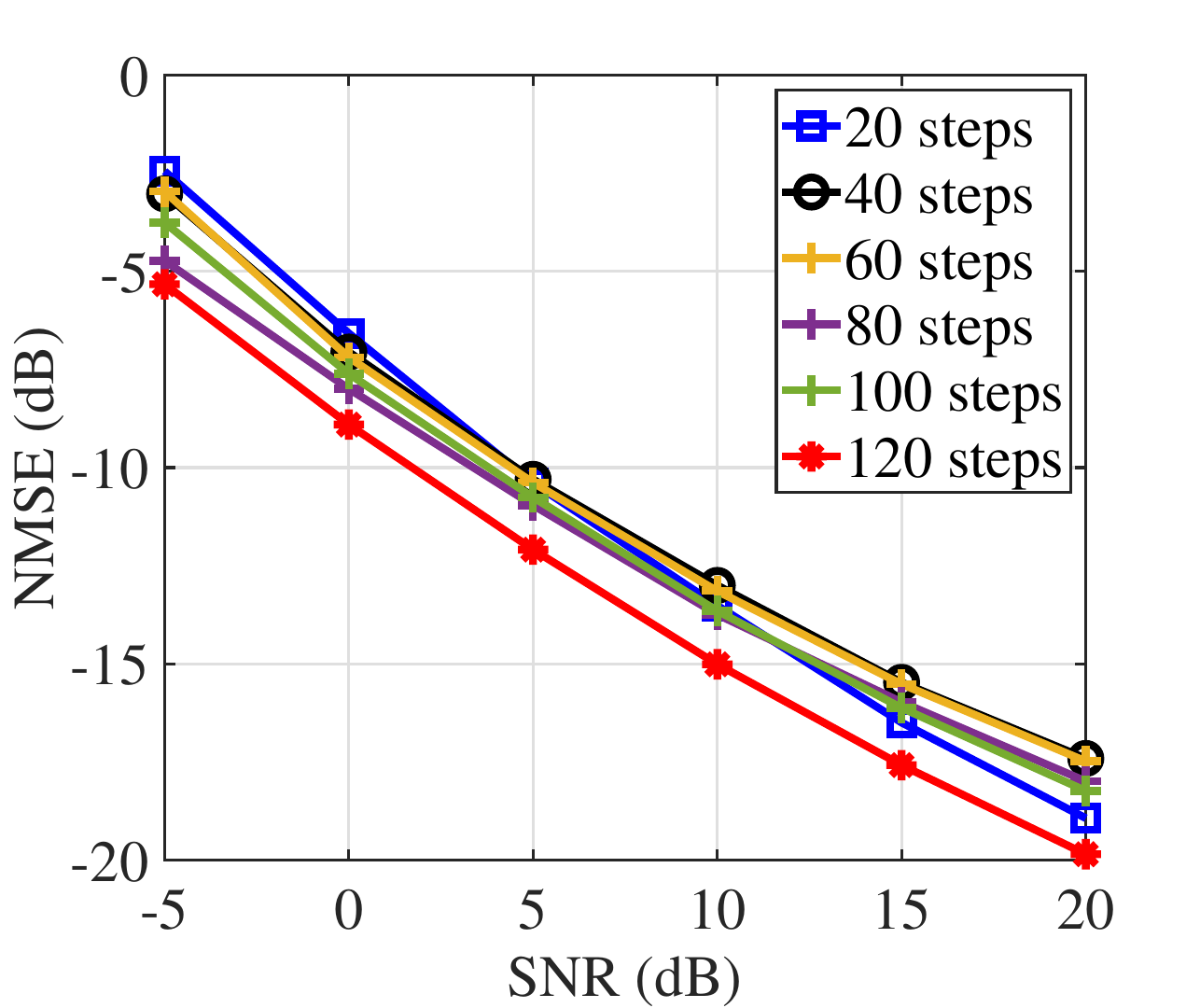}} 
    \subfigure[0.3 THz channel.]{ \includegraphics[width = 0.4\textwidth]{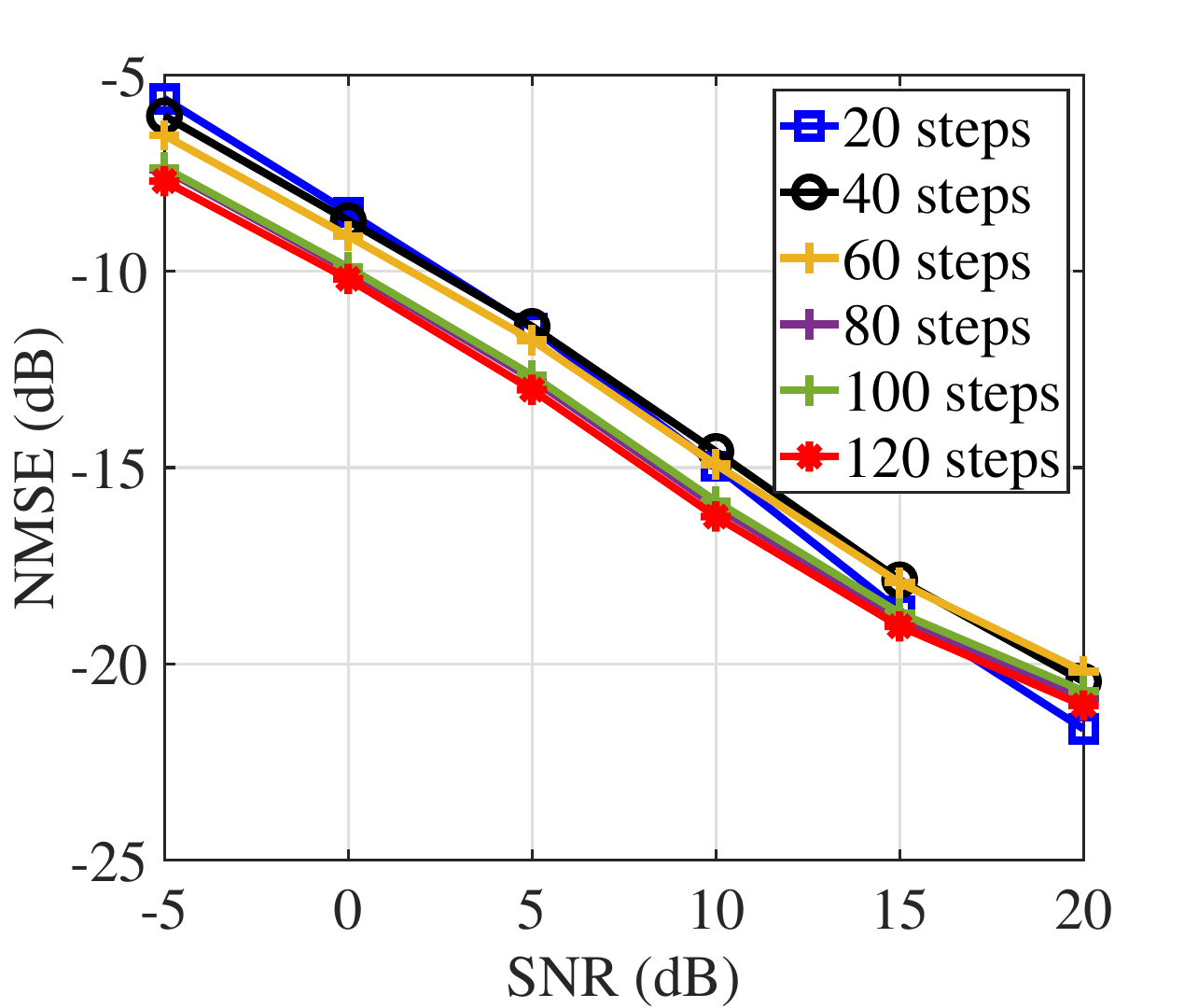}} 
    \caption{{Channel estimation accuracy of the DiffPace algorithm at $\alpha=0.8$ with different steps $K$.}}
    \label{fig_evaluation_step}
\end{figure*}

\subsection{Convergence Evaluation}
Fig.~\ref{fig_loss}(a) presents the convergence characteristics of the diffusion model through the evolution of the training and testing losses across successive epochs, where each epoch corresponds to a complete iteration through the training dataset. The testing loss curves for both mmWave and THz channels demonstrate rapid convergence within 40 epochs, confirming the model's capability to efficiently learn channel distribution patterns. Minor fluctuations observed in the training loss originate from the inherent randomness of stochastic gradient descent optimization, which the EMA strategy successfully mitigates. This stabilization effect becomes evident through the testing loss curves, which exhibit consistently decreasing trends in contrast to the training loss variations.

The practical deployment of CE often encounters limitations in the available training data. Fig.~\ref{fig_loss}(b) investigates the relationship between testing loss reduction and the proportion of utilized training samples. The testing error shows significant improvement until reaching 40\% data utilization, equivalent to approximately 4000 mmWave samples or 800 THz samples, beyond which the enhancement rate progressively stabilizes past the 60\% threshold. These results highlight the diffusion model's advantage, namely its robust feature extraction capability enables effective learning from limited samples. Such data-efficient characteristics make the DiffPace method particularly suitable for real-world implementations, where extensive channel measurement collections may be impractical.

\begin{figure*}[t]
    \centering
    \subfigure[60 GHz channel]{\includegraphics[width=0.4\textwidth]{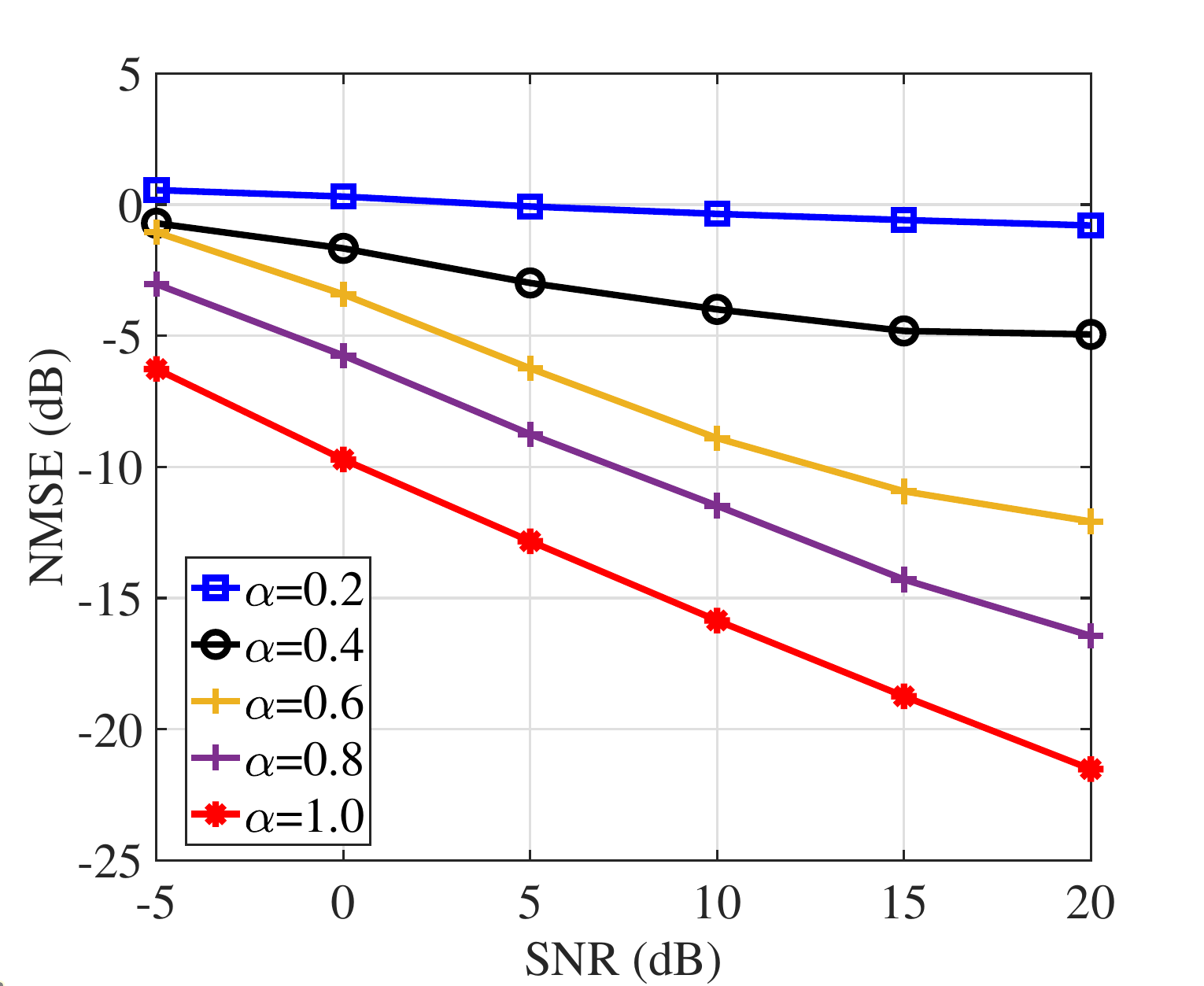}} 
    \subfigure[0.3 THz channel]{ \includegraphics[width = 0.4\textwidth]{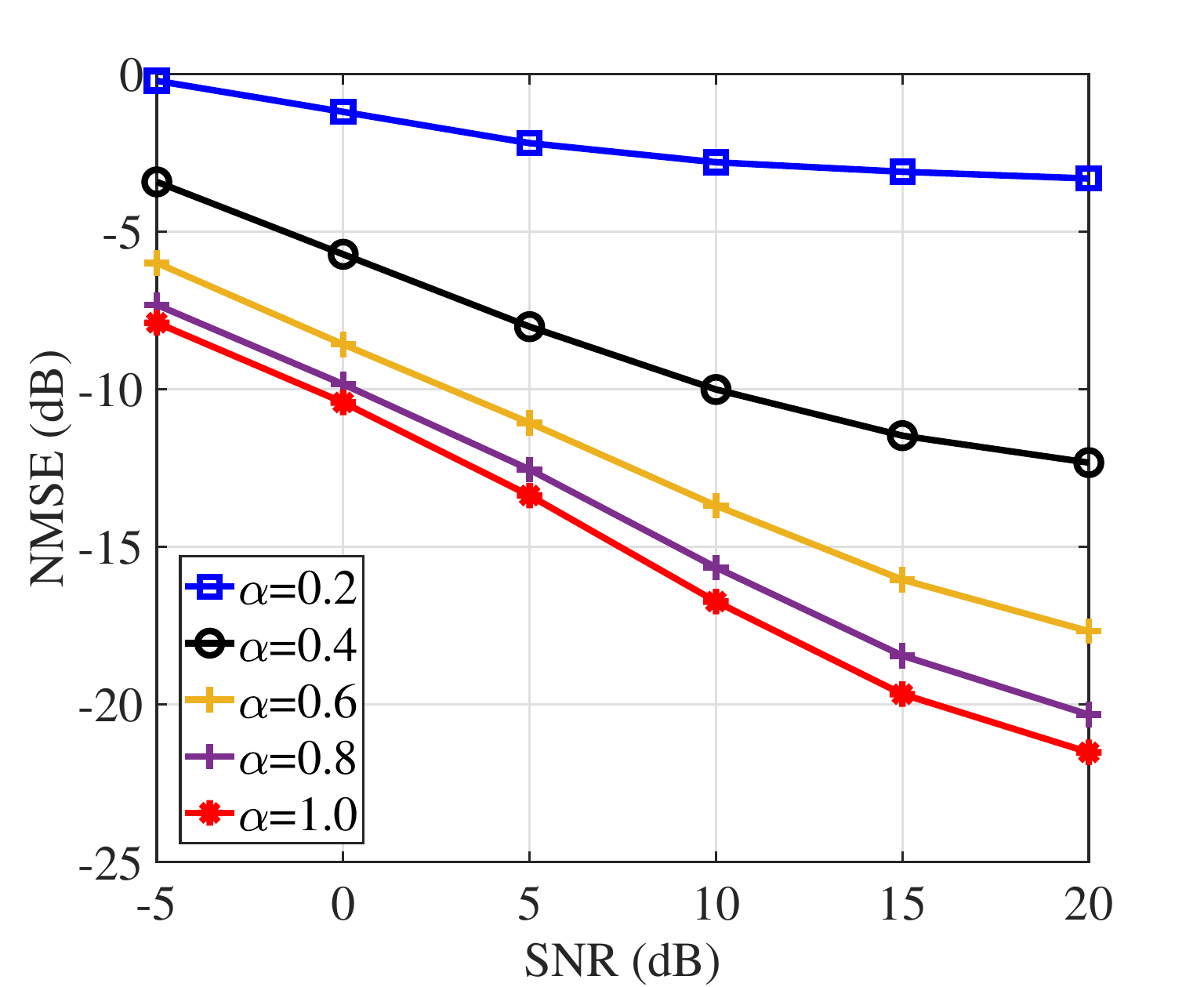}} 
    \caption{{CE accuracy of the DiffPace algorithm with $K=100$ and different pilot ratios.}}
    \label{fig_evaluation_alpha}
\end{figure*}

\subsection{Estimation Accuracy}
\revise{To evaluate the estimation accuracy of DiffPace, we compare it against several benchmarks: OMP~\cite{b14}, AMP~\cite{b13}, simultaneous OMP (SOMP)~\cite{omp11}, the minimum mean squared error (MMSE) estimator~\cite{b19}, the off-grid Sparse Bayesian Learning (SBL) method~\cite{SBL}, and learned denoising-based AMP (LDAMP)~\cite{b17}. The baseline methods OMP, AMP, SOMP, and LDAMP are implemented under the standard on-grid assumption, making them susceptible to performance degradation under off-grid conditions. In contrast, the off-grid SBL method incorporates adaptive grid refinement to partially alleviate basis mismatch. However, its performance remains constrained by the underlying parametric sparsity model and linear approximations of the channel structure. Moreover, SBL requires iterative expectation–maximization updates and matrix inversions in each iteration, leading to substantial computational overhead for large-scale antenna arrays. By comparison, DiffPace employs a data-driven generative prior learned via a diffusion model, which captures higher-order and non-Gaussian dependencies of hybrid near-/far-field channels. This enables DiffPace to handle off-grid effects more effectively and achieve superior reconstruction fidelity without relying on a predefined dictionary. For a fair comparison, all methods are evaluated on the same channel datasets, and the estimation results are presented in Fig.~\ref{fig_accuracy}.}

In Fig.~\ref{fig_accuracy}(a), the estimation accuracy of the benchmark methods are compared with the DiffPace method for the mmWave channel. \revise{The DiffPace method achieves the lowest estimation error among all considered approaches. Specifically, DiffPace attains an NMSE below –10 dB at a signal-to-noise ratio (SNR) of 10 dB, whereas other methods such as OMP, AMP, SOMP, and LDAMP fail to reach this level even at a higher SNR of 20 dB. This performance advantage arises from the inherent limitations of conventional approaches. Methods like OMP, AMP, SOMP, and LDAMP rely on fixed sparsifying dictionaries, making them vulnerable to basis mismatch in hybrid near-/far-field channels. The off-grid SBL method partially alleviates this issue through adaptive grid refinement, yielding improved results over the on-grid counterparts, but it remains inferior to DiffPace due to its dependence on parametric modeling assumptions. Although the MMSE estimator is theoretically optimal if the second-order channel statistics are perfectly known, such statistics are difficult to obtain in practice and fail to capture the hybrid near-/far-field characteristics of UM-MIMO channels.} In comparison, the proposed DiffPace is based on the HPSM model, which achieves better sparsifying performance compared to the traditional DFT codebook~\cite{hpsm_yuhan}. Moreover, the DiffPace utilizes the diffusion model to learn the distribution of channel. By the learned prior knowledge of the channel, the performance of CE is enhanced.

In Fig.~\ref{fig_accuracy}(b), the estimation accuracy of DiffPace for the THz channel is compared with other methods. DiffPace can still achieve the lowest NMSE, with -7 dB at an SNR of -5 dB, and can reach below -10 dB at SNR of 5 dB. This again shows the superiority of the DiffPace method in terms of CE accuracy compared with other methods.

When the results for the mmWave channel and THz channel are compared, for the THz channel, a lower MSE can be achieved, although more antennas are used for THz transmission. This is because the THz channel is sparser than the mmWave channel with only few propagation paths. From another point of view, the performance of DiffPace does not degrade due to the large dimension of the channel, but can benefit from the sparse nature of THz channels.

\subsection{Number of Inference Steps and Pilot Ratios}

Fig.~\ref{fig_evaluation_step} shows the relationship between CE accuracy and the number of inference steps. The observed performance improvement stems from the reduced step size in the ODE solver as the step count increases, which decreases approximation errors in the numerical integration. While the estimation accuracy shows gradual improvement with more steps, even the minimal 20-step configuration achieves respectable performance with -15 dB and -10 dB NMSE at 10 dB SNR for mmWave and THz systems respectively. However, it should be noted that the CE accuracy does not monotonically improve with increasing step counts across all SNR values. This occurs because the chosen parameters $\lambda$ and $\beta$, which govern the step length $\rho_i$ in~\eqref{eq_update2}, are not optimal for all SNR conditions. Notably, the performance degradation when reducing steps from 100 to 20 remains below 3 dB, demonstrating the ODE solver's effective acceleration capability while maintaining acceptable estimation quality.

The pilot ratio $\alpha$ similarly impacts estimation performance as shown in Fig.~\ref{fig_evaluation_alpha}. Performance improves monotonically as $\alpha$ increases from 0.2 to 1.0, with particularly significant gains occurring between 0.4 and 0.6. At $\alpha=0.6$ and 15 dB SNR, DiffPace achieves an NMSE below -10 dB for mmWave and -15 dB for THz systems, representing a relatively good performance. 

Interestingly, the THz system shows performance saturation at $\alpha=0.8$, while the mmWave system continues to benefit from additional pilots up to $\alpha=1.0$. This difference arises from the inherently sparser nature of the THz channel compared to the mmWave channel, allowing effective estimation with fewer measurements. The THz channel's greater compressibility enables satisfactory performance at lower pilot ratios than its mmWave counterpart.

\begin{figure}
    \centering
    \includegraphics[width=0.4\textwidth]{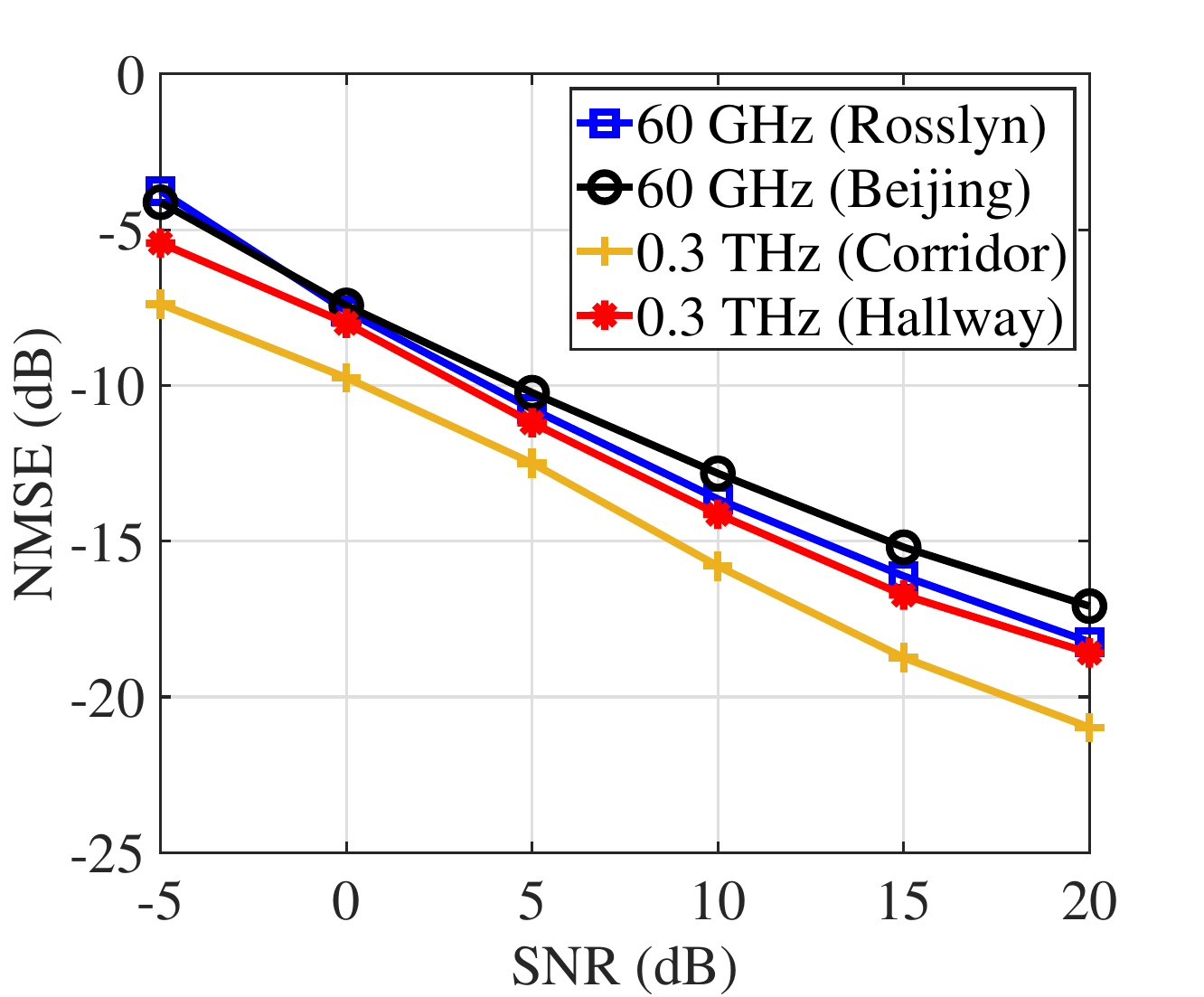}
    \caption{CE accuracy of the DiffPace algorithm in different environments.}
    \label{fig_evaluation_env}
\end{figure}
\subsection{Generalization Evaluation}
The generalization capability of DiffPace across diverse environments represents a critical requirement for practical deployment in mmWave and THz UM-MIMO systems. We evaluate this property by testing the trained model in scenarios that differ from its training environment, where performance degradation serves as our primary evaluation metric.

For mmWave systems, the model trained on the Rosslyn dataset (s002) maintains consistent estimation accuracy when tested in Beijing dataset (s007)~\cite{b23}, as evidenced by Fig.~\ref{fig_evaluation_env}. This minimal performance degradation demonstrates a significant advantage over supervised learning approaches since the diffusion model learns fundamental channel characteristics rather than overfitting to environment-specific features, enabling robust performance in new settings without requiring retraining.

In THz scenarios, DiffPace exhibits similar generalization capabilities. When trained on corridor environments and tested in hallway configurations~\cite{hall}, the method preserves high accuracy. This result confirms that the learned prior knowledge captures universal channel properties rather than environment-specific artifacts.

The plug-and-play architecture provides additional adaptability benefits. The diffusion model maintains fixed input dimensions corresponding to the channel size regardless of variations in the number of pilots. No structural modifications are required for different deployment configurations. Furthermore, the model automatically adapts to new environmental parameters through its learned priors. These characteristics make DiffPace particularly suitable for real-world deployment where environmental conditions may vary substantially. \revise{In addition, the same design principle provides inherent robustness to hardware-related impairments such as phase noise and non-linear distortion. Since these effects act as structured disturbances on the measurements, the learned generative prior helps suppress them by pulling estimates toward high-probability channel realizations. Moreover, robustness can be further improved by retraining or fine-tuning the diffusion prior with impairment-inclusive data, ensuring that the model captures the statistics of such distortions. This flexibility highlights DiffPace's potential for reliable operation under practical hardware constraints.}

\begin{table*}[t]
\caption{Comparison of computational complexity and execution time.}
\begin{center}
\begin{tabular}{lccc}
\toprule
\textbf{Method of CE} & \textbf{Complexity} & \textbf{60 GHz channel (seconds)} & \textbf{0.3 THz channel (seconds)} \\
\midrule
OMP~\cite{b14} & $\mathcal{O}(SN_tN_r)$ & 0.050 & 0.59 \\
AMP~\cite{b13} & $\mathcal{O}(KN_tN_r)$ & 0.050 & 0.62 \\
SOMP~\cite{omp11} & $\mathcal{O}(SN_tN_r)$ & 0.040 & 0.60 \\
MMSE~\cite{b19} & $\mathcal{O}(N_t^3N_r^3)$ & 0.10 & 5.56 \\
SBL~\cite{SBL} & $\mathcal{O}(KLN_t^3N_r^3)$ & 76.52 & 838.19 \\
LDAMP~\cite{b17} & $\mathcal{O}(N_tN_rK\sum_{l_c=1}^{L_c}F_{l_c}^2N_{l_c-1}N_{l_c})$ & 0.040 & 0.041 \\
DiffPace & $\mathcal{O}(N_tN_rK(M+\sum_{l_c=1}^{L_c}F_{l_c-1}N_{l_c}))$ & 0.032 & 0.032 \\
\bottomrule
\end{tabular}
\label{tab:complexity_runtime}
\end{center}
\end{table*}
\subsection{Computational Complexity}

Table~\ref{tab:complexity_runtime} compares the computational complexity of various channel estimation methods. The OMP and SOMP algorithm share the same order of complexity $\mathcal{O}(SN_tN_r)$, where $S$ represents channel sparsity and $N_t$, $N_r$ denote the number of transmit and receive antennas, respectively. Notably, the SOMP algorithm can achieve better estimation performance by exploiting common sparse support across subcarriers. The AMP algorithm demonstrates similar complexity $\mathcal{O}(KN_tN_r)$, with $K$ indicating the iteration count. In contrast, the MMSE estimator incurs significantly higher complexity at $\mathcal{O}(N_t^3N_r^3)$. \revise{Moreover, the computation complexity of the off-grid SBL method is $\mathcal{O}(KLN_t^3N_r^3)$, where $K$ represent the number of iterations and $L$ represents the number of grid refinement steps.}

For DL approaches, the complexity of the CNN network depends on its architecture, calculated as $\mathcal{O}(N_tN_r\sum_{l_c=1}^{L_c}F_l^2N_{l_c-1}N_{l_c})$, where $l_c=1,\cdots,L_c$ indexes the CV layer, $F_{l_c}$ denotes the filter size, and $N_{l_c}$ is the feature map count of layer $l_c$. Both LDAMP and DiffPace employ CNN networks, with LDAMP requiring $\mathcal{O}(N_tN_rK\sum_{l_c=1}^{L_c}F_{l_c}^2N_{l_c-1}N_{l_c})$ operations for $K$ iterations. DiffPace introduces an additional complexity of $\mathcal{O}(N_tN_rM)$ due to its operation of measurement error projection onto the learned channel manifold, where $M$ represents the number of training slots. The overall complexity of DiffPace is $\mathcal{O}(N_tN_rK(M + \sum_{l=1}^{l_c}F_l^2N_{l_c-1}N_{l_c}))$. The CNN used in DiffPace contains approximately 210,372 trainable parameters. For a single inference, the computational cost amounts to 0.39 giga floating-point operations (GFLOPs) for mmWave channel input and 6.44 GFLOPs for THz channel input. \revise{While DiffPace exhibits marginally higher complexity than other methods (except MMSE and SBL), it achieves superior estimation accuracy. The computational complexity of DiffPace grows linearly with $N_tN_r$, making it scalable to large antenna arrays. In our experiments with $N_r=64$ and $N_t=256$, DiffPace remained computationally tractable. To support scalability, we adopted a lightweight CNN backbone, since alternatives such as transformer-based models incur quadratic complexity in the channel dimension $N_tN_r$, which becomes impractical for ultra-large arrays.}

\revise{To further assess computational efficiency, we also compare the actual execution time of all methods, as summarized in Table~II. For a fair comparison, the number of iterations or inference steps is fixed to 20 for all methods except MMSE, since the execution time scales approximately linearly with the number of steps. For the mmWave channel, the runtimes of OMP, AMP, SOMP, MMSE, SBL, LDAMP, and DiffPace are 0.050, 0.050, 0.040, 0.10, 76.52, 0.040, and 0.032 seconds, respectively. For the THz channel, the corresponding runtimes are 0.59, 0.62, 0.60, 838.19, 5.56, 0.041, and 0.032 seconds. For the mmWave dataset, the latency of OMP, AMP, and SOMP is comparable to that of DiffPace. However, in the THz case, their runtime grows linearly with the channel dimension $N_rN_t$, whereas both LDAMP and DiffPace maintain nearly constant latency. The MMSE and SBL methods incur even higher runtimes, since their complexities scale as $\mathcal{O}(N_r^3N_t^3)$, which become prohibitive in the THz regime. The advantage of LDAMP and DiffPace lies in their ability to leverage the strong parallelism of GPU computing, allowing them to sustain stable latency even for large-scale arrays. Furthermore, DiffPace adopts a lightweight CNN backbone, while LDAMP employs a deeper CNN architecture with more convolutional layers, resulting in a higher latency for LDAMP. Consequently, DiffPace achieves the lowest runtime among the considered methods while retaining scalability across both datasets.}

\section{Conclusion}\label{sec:conclusion}
In this paper, we propose DiffPace, a novel framework for CE in mmWave and THz MIMO systems based on the HPSM channel model. By incorporating a diffusion model as a generative prior within a plug-and-play architecture, DiffPace effectively balances the prior learned by the diffusion network with measurement consistency, enabling accurate channel recovery. This ensures the estimated channel matches the target distribution while remaining consistent with the received signal. Additionally, deterministic sampling using ODE solvers reduces the number of diffusion steps by over 90\% compared with stochastic sampling, from more than 1000 to fewer than 100, while maintaining competitive accuracy. Experimental results demonstrate that DiffPace outperforms existing methods in cross-field scenarios, achieving a 5 dB improvement in NMSE over state-of-the-art methods.

Looking ahead, two promising directions emerge to further enhance efficiency. First, integrating consistency models can enable single-step or few-step sampling by enforcing output alignment across time steps, building on the acceleration principles explored here. Second, network compression techniques could compress the diffusion prior into a lightweight model, reducing computational overhead without sacrificing accuracy. These advancements would align with the growing demand for real-time CE in ultra-broadband mmWave and THz systems, paving the way for practical deployment in 6G networks. By bridging the gap between iterative refinement and rapid inference, DiffPace establishes a flexible foundation for future work on physics-aware generative models in high-frequency communications.

\bibliographystyle{IEEEtran}
\bibliography{main}

\end{document}